\documentclass[prb,a4paper,showpacs,twocolumn]{revtex4}

\usepackage{amssymb}
\usepackage{amsmath}
\usepackage{amsfonts}
\usepackage{graphicx}
\usepackage{epstopdf}
\usepackage{bm}
\usepackage{color}
\renewcommand{\Re}{\,\textrm{Re}\,}
\renewcommand{\Im}{\,\textrm{Im}\,}
 \DeclareMathOperator{\tr}{tr}
\DeclareMathOperator{\sign}{sgn} \DeclareMathOperator{\Imag}{Im}

\begin{document}

\title{The problem of ``macroscopic charge quantization" in
single-electron
devices
}

\author{I.S.\,Burmistrov$^{1,2}$ and A.M.M.\,Pruisken$^{3}$}
\affiliation{$^{1}$ L.D. Landau Institute for Theoretical Physics,
Kosygina street 2, 117940 Moscow, Russia} \affiliation{$^{2}$ Department
of Theoretical Physics, Moscow Institute of Physics and
Technology, 141700 Moscow, Russia}
\affiliation{$^{3}$Institute for Theoretical Physics, University of Amsterdam,
Valckenierstraat 65, 1018XE Amsterdam, The Netherlands}

\begin{abstract}
In a recent Letter by the authors [I.S.\,Burmistrov and A.M.M.\,Pruisken, Phys. Rev. Lett. 101, 056801 (2008)] 
it was shown that
single-electron
devices (single electron transistor or SET) display ``macroscopic
charge quantization" which is completely analogous to the quantum
Hall effect observed on very different electron systems. In this
investigation we present more detail on these new findings. Based
on the Ambegaokar-Eckern-Sch\"{o}n (AES) theory of the Coulomb
blockade we introduce a general response theory that probes the
sensitivity of SET to changes in the boundary conditions. This
response theory defines a new set of physical observables and we
establish the contact with the standard results obtained from
ordinary linear response theory. The response parameters generally
define the renormalization behavior of the SET in the entire
regime from weak coupling with large values of the tunneling
conductance all the way down to the strong coupling phase where
the system displays the Coulomb blockade. We introduce a general
criterion for charge quantization that is analogous to the
Thouless criterion for Anderson localization. We present the
results of detailed computations on the weak coupling side of the
theory, i.e. both perturbative and non-perturbative (instantons).
Based on an effective theory in terms of quantum spins we study
the quantum critical behavior of the AES model on the strong
coupling side. Consequently, a unifying scaling diagram of the SET
is obtained. This diagram displays all the super universal
topological features of the $\theta$ angle concept that
previously arose in the theory of the quantum Hall effect.
\end{abstract}


\maketitle


\section{Introduction\label{Sec:Intro}}
\subsection{Coulomb blockade}
The Coulomb blockade in nanostructures is one of the cornerstones
of modern condensed matter physics. The simplest approach to
electron tunneling through quantum dots was proposed by
Ambegaokar, Eckern and Sch\"{o}n in $1982$.~\cite{AES,SZ} 
Their model (in brief AES model) became the focus of a stream of experimental and
theoretical papers~\cite{SET,ZPhys,Glazman} following the first experimental indications of ``macroscopic charge quantization'' in single electron devices in $1991$.~\cite{QExpFirst}

To experimentally control the transport of electrons~\cite{Exp1Exp2} one generally uses the so-called ``single electron transistor" or SET.~\cite{SETExp} This is a mesoscopic metallic island that is capacitively coupled to a gate
and connected to two metallic reservoirs through tunneling contacts
with a total conductance $g$ (see Fig.~\ref{Figure1}a).

The experimental conditions of the AES model are limited and extremely 
well known.~\cite{Falci,BEAH,ET,Glazman} The model is
nevertheless richly complex and much of the physical consequences have
remained unknown. Over the years, however, it has
slowly become more evident that the AES theory of the Coulomb blockade 
is in many ways similar to the theory of the quantum Hall effect.~\cite{PruiskenBurmistrov2} For example, the AES model is asymptotically 
free in $0+1$ space-time dimension, possesses instantons and has an 
instanton angle $\theta$. This immediately raises the fundamental 
question whether the experimental phenomenon of ``macroscopic charge 
quantization" in the SET is possibly related to the 
``robust quantization" of the Hall conductance observed on very different 
electronic systems.

The AES model has a number of very significant advantages as compared to the more conventional theories of the $\theta$ vacuum or instanton vacuum. For example, the winding numbers of the theory (``topological charge") are quantized at the outset of the problem. This is quite unlike the usual situation where the historical controversies in quantum field theory continue to haunt the subject. For example, it has been pointed out only very recently that the $\theta$ vacuum concept generally displays ``massless chiral edge excitations" that are very different from those in the ``bulk" of the system. Disentwining these different types of excitation is synonymous for separating the {\em fractional} topological sectors of the theory from the {\em integral} ones. 

Remarkably, it turns out the existence of ``massless chiral edge excitations"
in the problem automatically reveals the existence of the quantum Hall effect. 
This fundamental phenomenon previously remained concealed. However, it provides the resolution to longstanding problems such as the quantization of topological charge, the meaning of instantons and instanton gases etc. etc. 

The existence of ``massless chiral edge excitations" has furthermore led to the idea of ``super universality" which states that all the fundamental features of the instanton vacuum concept are precisely those of the quantum Hall effect.~\cite{PruiskenBurmistrov2} These include not only the robust quantization of the Hall conductance but also the existence of ``gapless excitations" at $\theta =\pi$ or, equivalently, ``quantum criticality" of the quantum Hall plateau transition.

It is of interest to know whether these new advances possibly also apply to
the AES model. In this case, the microscopic origins of the integral and fractional topological sectors are far more obvious. For example,
the integral sectors directly emerge from quantum statistics and they describe 
the quantum system (SET) in thermal equilibrium. On the other hand, the fractional topological sectors do not describe ``edge" excitations but, rather, they have the meaning of perturbing external fields that take the SET out of thermal equilibrium. 
The great advantage of the AES model, however, is that the $\theta$ dependence can be studied on the strong coupling side. The AES model is therefore an outstanding laboratory where the various different aspects of ``super universality" can be explored and investigated in great detail.

It should be mentioned that the AES model in a different context is also known 
as the ``circular brane model.''~\cite{LZ} It is furthermore of direct 
physical interest in the theory of granular metals at intermediate
temperatures.~\cite{Cond}
\subsection{Charge quantization}
The phrase ``macroscopic charge quantization" usually refers to the charge
of an isolated {island} that is disconnected from the reservoirs. 
It is given by the naive strong coupling limit of the AES model where the tunneling
conductance $g$ is put equal to zero.

{This naive} approach leads to the electrostatic picture of the Coulomb blockade where the average charge ($Q$) on the island is {\em robustly} 
quantized in units of $e$ as the temperature ($T$) goes to absolute zero. 
This quantization breaks down for very special values of the gate voltage 
\begin{equation}
 V_g^{(k)}=e(k+1/2)/C_g
\end{equation}
where $k$ is an integer and $C_g$ denotes the gate capacitance. At these very special values of $V_g$ a first order {\em quantum phase transition} occurs separating two different phases with $Q=k$ and $k+1$ respectively.

The electrostatic picture of the Coulomb blockade gets fundamentally
complicated when the tunneling conductance $g$ is finite.
It is well known, for example, that
due to the strong charge fluctuations in the SET the averaged
charge $Q$ on the {island} is generally {\em un}-quantized.~\cite{Matveev} 
Despite the impressive list of existing theoretical work
on both the strong coupling side~\cite{Glazman,Schon} 
($g\ll 1$) and weak coupling side~\cite{Cond} ($g\gg 1$) of the problem 
it is not known what the electrostatic or semi classical picture of the SET 
exactly stands for. This fundamental drawback clearly upsets the 
concept of ``robust charge quantization" in single electron 
devices.
\subsection{Outline of this investigation}
The main objective of this investigation is to show that the SET  
displays macroscopic charge quantization in much the same way
as the two dimensional electron gas displays the quantum Hall effect.
We develop a complete quantum theory of the SET and 
introduce a unifying scaling diagram that spans the entire range 
from weak to strong coupling.

We benefit from the advances made over the years in the theory of 
the quantum Hall effect. We introduce, in particular, the 
``physical observables" of the AES theory that measure the
sensitivity of the SET to changes in the boundary conditions.
In the present context this means that the quantum system is 
taken out of thermal equilibrium by perturbing fields. The main 
problem to be solved is how to lay the bridge between the 
sensitivity to the boundary conditions on the one hand, and 
the standard expressions for linear response obtained from 
the Kubo formalism on the other.
\subsubsection{Electrostatic picture revisited}
To start we briefly review the microscopic origins of the AES
model and summarize the results known from previous work in
Section \ref{Sec:AES}. 

To see the concept of ``physical observables" at work   
we consider in Section \ref{Sec:Phys} 
the trivial case of an isolated island at finite $T$ obtained by 
putting the tunneling conductance $g$ equal to zero. 
This simple but instructive example sets the stage for most of the analysis in 
the remainder of this paper. 

We point out, first of all, that the averaged charge $Q$ on the isolated island 
is a measure of the sensitivity of the system to changes in the boundary conditions. This notion immediately suggests a generalized Thouless criterion that relates the robust quantization of $Q$ on the island to the appearance of an energy gap.

Secondly, we show how the renormalization behavior of $Q$ at finite $T$ provides a complete knowledge of the low energy dynamics of the isolated island. This behavior involves two different kinds of fixed points, i.e. stable ones at $Q=k$ which describe the robust quantization of charge as $T$ goes to zero, and unstable 
ones at $Q=k+1/2$ describing the transition between the states $Q=k$ and $k+1$ of the island.
\subsubsection{Two sets of physical observables}
Armed with the insights obtained from the isolated island we next 
embark on the general problem with finite $g$ in Section ~\ref{Kubo}. 
We introduce two slightly different but physically equivalent sets 
of response parameters $g^\prime$ and $q^\prime$. The different 
expressions that we obtain stem from slightly different ways of handling the 
fractional topological sectors of the AES theory. 

The first set is the simpler one which is a direct generalization of the results obtained for an isolated island. The second set is slightly more involved but our findings permit a direct comparison with the expressions obtained from linear response theory.

In both cases, however, one may think of $g^\prime$ and $q^\prime$ in terms of the 
sensitivity of the SET to changes in the boundary conditions. In both cases also one may think of $g^\prime$ in terms of the SET conductance. The quantity $q^\prime$ is new and in general very different from the conventionally studied averaged charge 
$Q$ on the island. Within linear response theory we express $q^\prime$ in terms of the antisymmetric current-current correlation function. We identify this new quantity with the previously unrecognized quasi particle charge of the SET.

In complete analogy with the theory of the quantum Hall effect, we 
relate the conditions for ``macroscopic charge quantization" 
$g^\prime = 0$ and $q^\prime = k$ with integer $k$ 
to the appearance of an energy gap in the SET. 
For $g=0$ these conditions are identically the same as those
obtained from the electrostatic picture of the SET. For finite $g$, 
however, these conditions describe an entirely different physical 
state of the SET. They describe the macroscopic quantization of the
{\em quasi particle charge}, rather than  the {\em averaged charge} 
on the island.
\subsubsection{Explicit computations}
This takes us to the second part of this investigation where
we explicitly compute, in Sections ~\ref{Sec:Weak} and ~\ref{Sec:Strong}, 
the observable theory $g^\prime$ and $q^\prime$ 
in the various different regimes in $g$ of interest. 
We benefit from having two different definitions of 
$g^\prime$ and $q^\prime$. The different computational schemes provide 
a direct check on the universal and non-universal parts of the AES theory.    

In Section~\ref{Sec:Weak} we consider the weak coupling phase 
$g \rightarrow \infty$ of the AES model. We report the detailed 
results for the renormalization group $\beta$ functions based on ordinary perturbation theory as well as instantons. 
Even though this Section is self-contained, we refer the reader to
the literature for a more detailed exposure to the instanton calculational 
technique.~\cite{PruiskenBurmistrov2}

In Section~\ref{Sec:Strong} we address the strong coupling phase of the AES theory and study, in particular, the quantum critical behavior of the SET at finite $g$. For this purpose we first map the critical behavior of the AES model onto an effective theory of quantum spins. We employ Abrikosov's pseudo fermion technique and extract the $\beta$ functions of the AES theory near the critical point. 

The most important results of this investigation are encapsulated in the unifying scaling diagram in the $g^\prime$ - $q^\prime$ plane as illustrated in Fig. \ref{RGFIG}. The flow lines clearly indicate that the phenomenon of ``macroscopic charge quantization" is a universal feature of single electron devices that always appears in the limit where $T$ goes to zero. Fig. \ref{RGFIG} furthermore displays 
all the super universal features of the $\theta$ angle concept that previously arose in the theory of the quantum Hall effect.

We end the paper with a conclusion in Section~\ref{Sec:End}.


\section{AES model\label{Sec:AES}}

\begin{figure}[b]
\includegraphics[height=75mm]{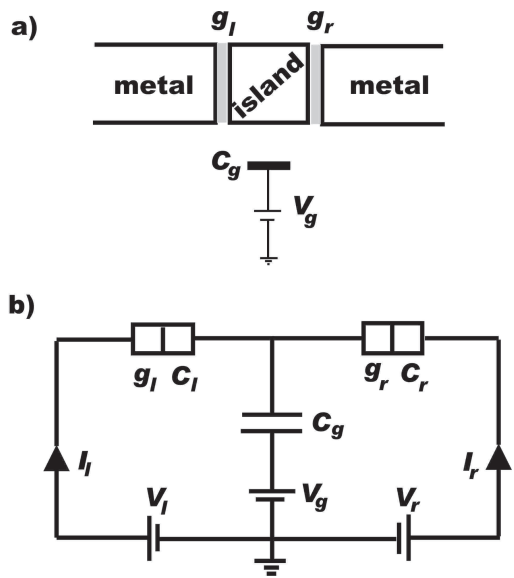}
\caption{a) Sketch of the SET device. b) Equivalent circuit of the
SET.} \label{Figure1}
\end{figure}

\subsection{Action}

It is well understood by now that the AES model of the Coulomb
blockade is a limiting case of the so-called universal theory of zero
dimensional electron systems.~\cite{UnivHam} 
To start we briefly review the
microscopic origins of this model. The experimental
design of the SET is illustrated in Fig.~\ref{Figure1}a. The
hamiltonian is split in three distinctly different parts
\begin{gather}
\mathcal{H} = \mathcal{H}_0 +  \mathcal{H}_c + 
\sum_{s=l,r} \mathcal{H}_T^{(s)}. \label{HamStart}
\end{gather}
The first part is the free electron piece 
\begin{gather}
\mathcal{H}_0 = \sum_{k,s=l,r} \epsilon^{(s)}_{k}
a^{(s)\dag}_{k} a^{(s)}_{k} + \sum_{\alpha} \epsilon_{\alpha}
d^\dag_{\alpha} d_{\alpha} .
\end{gather}
The index $s$ runs over of the reservoirs on the left hand
side ($l$) and right hand side ($r$) of the
{island} respectively. The subscript $k$ denotes the electronic states in the reservoirs and $\alpha$ those on the island. The 
$\epsilon_k^{(a)},\epsilon_\alpha$ are the energies relative to the Fermi level.

The second term in Eq. \eqref{HamStart} is the result of the Coulomb interaction between the electrons on the {island}
\begin{gather}
\mathcal{H}_c = E_c \left (\sum_{\alpha} d^\dag_{\alpha} d_{\alpha} - q \right
)^2. \label{HamCoul}
\end{gather}
$E_c = e^2/(2 (C_l+C_r+C_g))$ stands for the charging energy and
$q=C_g V_g/e$ represents the external charge on the{island}
(see Fig.~\ref{Figure1}b). 

The last part of Eq. \eqref{HamStart} describes the tunneling
of electrons between the reservoir and the island
\begin{gather}
\mathcal{H}_T^{(s)} = \sum_{k\alpha}
t_{k\alpha}^{(s)} a^{(s)\dag}_{k} d_{\alpha} +
\textrm{h.c.}
\end{gather}
The matrix $t_{k\alpha}^{(s)}$ contains the amplitudes for tunneling between the reservoirs and the island. To characterize this tunneling it is convenient to introduce the following hermitean matrices
\begin{gather}
   \hat{g}^{(s)}_{kk^\prime}=4\pi^2
   \left[\delta(\epsilon^{(s)}_{k})\delta(\epsilon^{(s)}_{k^\prime})\right]^{1/2}\sum_\alpha t^{(s)}_{k\alpha}
   \delta(\epsilon_{\alpha})t^{(s)\dagger}_{\alpha
   k^\prime}, \label{gm1}\\
   \check{g}^{(s)}_{\alpha\alpha^\prime}=4\pi^2
   \left[\delta(\epsilon_{\alpha})\delta(\epsilon_{\alpha^\prime})
   \right]^{1/2}\sum_k t^{(s)\dagger}_{\alpha
   k}
   \delta(\epsilon^{(s)}_{k})t^{(s)}_{k\alpha^\prime}.\label{gm2}
\end{gather}
The first matrix acts in the Hilbert space of states of a single
reservoir and the second one in the Hilbert space of states of the island. 
One should think of the delta-functions in Eqs.~\eqref{gm1}-\eqref{gm2} as being  smoothed out over a scale $\delta E$ such that $\max\{\delta,\delta^{(l,r)}\} \ll\delta E\ll T$. Here, $\delta$ and $\delta^{(l,r)}$ stand for mean level spacing of single-particle states on the island and reservoirs respectively.

The classical dimensionless conductance (in units $e^2/h$) of the junction between a reservoir and the island can be  expressed as follows~\cite{Glazman}
\begin{gather}
 \label{conductance-def1}
 g_s=\sum_k \hat{g}^{(s)}_{kk} \equiv \sum_\alpha\check{g}^{(s)}_{\alpha\alpha} .
\end{gather}
Therefore, each non-zero eigenvalue of $\hat g^{(s)}$ or $\check g^{(s)}$ corresponds to the transmittance of some `transport' channel between a reservoir and the island.~\cite{Landauer} The effective number of these `transport' channels 
($N^{(s)}_{\rm ch}$) is given by
\begin{equation}
N^{(s)}_{\rm ch} = \frac{\left ( \sum\limits_k \hat{g}^{(s)}_{kk} \right )^2}{\sum\limits_{kk^\prime} \hat g^{(s)}_{kk^\prime}\hat g^{(s)}_{k^\prime k}} \equiv \frac{\left ( \sum\limits_\alpha \check{g}^{(s)}_{\alpha\alpha} \right )^2}{\sum\limits_{\alpha\alpha^\prime} \hat g^{(s)}_{\alpha\alpha^\prime}\hat g^{(s)}_{\alpha^\prime \alpha}} .
\end{equation}
The effective dimensionless conductance $g^{(s)}_{\rm ch}$ of a `transport' channel can be written as follows
\begin{gather}
\label{aes-condition1}
g^{(s)}_{\rm ch}=  \frac{\sum\limits_{kk^\prime} \hat g^{(s)}_{kk^\prime}\hat g^{(s)}_{k^\prime k}}{\sum\limits_k \hat{g}^{(s)}_{kk} } \equiv \frac{\sum\limits_{\alpha\alpha^\prime} \hat g^{(s)}_{\alpha\alpha^\prime}\hat g^{(s)}_{\alpha^\prime \alpha}}{\sum\limits_\alpha \check{g}^{(s)}_{\alpha\alpha}} .
\end{gather}
The dimensionless conductance $g_s$ then becomes
\begin{equation}
g_s = g^{(s)}_{\rm ch}N^{(s)}_{\rm ch} .
\end{equation}
In what follows we will always assume 
\begin{gather}
\label{aes-condition2}
g^{(l,r)}_{\rm ch}\ll 1 .
\end{gather}
Notice that under these circumstances the conductances $g_{l,r}$ can still
be large provided the effective number of channels $N^{(l,r)}_{\rm ch}\gg 1$ is
sufficiently large.

We furthermore assume that the mean level spacing is negligible
$\delta \ll T/ \max\{1,g\}$,~\cite{BEAH} and the 
charging energy is sufficiently large $E_c\gg \delta$ such that the 
effects of the exchange interaction can be ignored.~\cite{Glazman,UnivHam}

\subsubsection{Path integral representation\label{Path}}
Given this sequence of limitations one can express the dynamics of
the SET in terms of a single abelian phase $\Phi(\tau)$ with
$\tau$ standing for the imaginary time.~\cite{AES} This field
generally describes the potential fluctuations on the island
according to $V(\tau)=i\dot{\Phi}(\tau)$.  The quantum mechanical
partition function $Z$ can be written as a sum over winding
numbers $W$ according to
\begin{equation} \label{Z-W}
 Z [q] = \sum_{W=-\infty}^{\infty} e^{2\pi i q W} Z_W
\end{equation}
where $Z_W$ is the integral over all paths $\Phi (\tau)$ that
start with $\Phi(0)$ at $\tau=0$ and end with $\Phi(\beta) = \Phi
(0) + 2 \pi W$ at $\tau =\beta$ with $\beta$ the inverse
temperature:
\begin{eqnarray} \label{Znlnr-n}
 Z_W &=& \int_{
 \Phi (\beta) = \Phi (0) + 2\pi W }
 \mathcal{D} \Phi(\tau)\, e^{-\mathcal{S}_d [\Phi]
 - \mathcal{S}_c [\Phi] } .
\end{eqnarray}
Here, the action $\mathcal{S}_d$ describes the tunneling between
the island and the reservoirs
\begin{equation} \label{SdStart}
 \mathcal{S}_d [\Phi] = \frac{g}{4}\int_{0}^{\beta} d\tau_1
 d\tau_2 ~
 \alpha(\tau_{12}) e^{i\Phi(\tau_1) - i\Phi(\tau_2)}
\end{equation}
where $g=g_l+g_r$ and $\tau_{12} = \tau_1-\tau_2$. The kernel
$\alpha(\tau) = \alpha (\tau+\beta)$ in time and frequency
representation is given by
\begin{equation}\label{alpha-tau}
\alpha(\tau) = - {T^2}{\textrm{cosec}^2 (\pi T \tau)} =
\frac{T}{\pi}\sum_{\omega_n}|\omega_n|e^{-i\omega_n\tau}
\end{equation}
with $\omega_n=2\pi T n$.
%
%
%
The second term $\mathcal{S}_c$ corresponds to the charging energy
due to the Coulomb interaction between the electrons on the island
\begin{equation}\label{ScStart}
 \mathcal{S}_c [\Phi] = \frac{1}{4 E_c} \int_{0}^{\beta} d \tau\,
 \dot{{\Phi}}^2.
\end{equation}
Finally, the exponential factor containing the external charge $q$
in Eq. \eqref{Z-W} describes the coupling between the island and
the gate of the SET.
\subsubsection{Functional integral representation}
\begin{figure}[b]
\includegraphics[width=60mm]{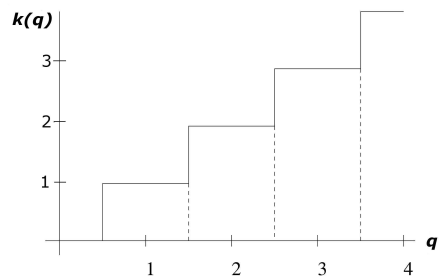}
\caption{Integer $k(q)$ and fractional $\theta(q)$ pieces of q.} \label{FIG_KT}
\end{figure}

An elegant formulation of the AES theory is obtained using the
$O(2)$ field variable
\begin{equation}
 \mathcal{Q} (\tau) = \left(
\begin{matrix} \cos \Phi & ~~\sin \Phi \\ \sin \Phi & -\cos \Phi \end{matrix}
\right)
\end{equation}
with $\mathcal{Q}^2 (\tau) = \bm{1}$. The partition function can now be
expressed as follows
\begin{equation}\label{Zstart-1}
 Z [q] = \int_{\partial V} \mathcal{D}[\mathcal{Q}] e^{-\mathcal{S} [\mathcal{Q}]}
\end{equation}
where the subscript $\partial V$ indicates that the functional
integral is performed with periodic boundary conditions
\begin{equation}\label{PBCs}
 \mathcal{Q}(0) = \mathcal{Q}(\beta) .
\end{equation}
The action is given by
%
\begin{eqnarray}\label{SdStart-1}
 \mathcal{S}[\mathcal{Q}] &=& \int_{0}^{\beta} d\tau_1
 d\tau_2 \gamma (\tau_{12}) \tr \mathcal{Q}(\tau_1) \mathcal{Q}(\tau_2) \notag \\
 &-&
 \frac{q}{2} \int_{0}^{\beta} d\tau \tr \sigma_y \mathcal{Q} \partial_\tau\mathcal{Q}
\end{eqnarray}
%
with $\bm{\sigma}$ denoting the Pauli matrices. The kernel
$\gamma(\tau_{12})$ in frequency representation can be written as
follows
\begin{equation}
 \gamma (i\omega_n) =  \frac{g}{4\pi} |\omega_n| + \frac{1}{8 E_c}
 \omega_n^2 .
\end{equation}
Alternatively one may express the action in terms of the $O(2)$
vector field, $\mathcal{Q}(\tau) = \sigma_z \left( n_x (\tau) + i\sigma_y
n_y (\tau) \right)$, with $\bm{n}^2 (\tau) = 1$.
The integer valued topological charge of the system can be
expressed in three different ways
\begin{eqnarray} \label{top-charge}
 \mathcal{C} [\mathcal{Q}] &=& -\frac{i}{4\pi} \int_0^\beta d \tau \tr \sigma_y \mathcal{Q} \partial_\tau \mathcal{Q}
 \nonumber \\
 &=& - \frac{1}{2\pi} \int_0^\beta d \tau
 \epsilon_{\mu\nu} \bm{n}_\mu \cdot \partial_\tau \bm{n}_\nu
 \nonumber \\
 &=& ~~\frac{1}{2\pi} \int_0^\beta d \tau\, \dot{\Phi}.
\end{eqnarray}
This quantity is nothing but the number of times ($W$) the $O(2)$
vector field $\bm{n}$ is winding around. It is important to
emphasize that both the periodicity statement of Eq. \eqref{PBCs}
and the quantization of topological charge in Eq.
\eqref{top-charge} are fundamental features of the AES theory that
inherently describe the SET in thermal equilibrium. This theory
only depends on the external charge $q$ modulo $2\pi$. If, for
example, one splits $q$ into a fractional piece $-\pi < \theta (q)
\le \pi$ and an integral piece $k(q)$ (see Fig.~\ref{FIG_KT})
\begin{equation}\label{split}
 q = \frac{\theta (q)}{2\pi} + k(q)
\end{equation}
then the quantum mechanical partition function only depends on the
fractional piece $\theta(q)$
\begin{equation}
 Z[q] = Z [\theta (q)/2\pi].
\end{equation}
To extract the integral piece $k(q)$ from the AES theory one must
in general consider perturbing fields that take the SET out of
thermal equilibrium.

%
\subsection{Instantons \label{Instantons}}
One of the most impressive features of the tunneling term of
Eq.~\eqref{SdStart} is that it possesses stable classical minima
$\Phi_W (\tau)$ for each topological sector $W$. We term these
classical solutions ``instantons" since they are completely
analogous to Yang-Mills instantons.~\cite{BPST} The general
expression for $\Phi_W (\tau)$ is given
by~\cite{Instantons1,Instantons11}
\begin{equation}
 e^{i\Phi_{W}(\tau)} = \prod \limits_{a=1}^{|W|}\frac{1- z(\tau) z_a}{z (\tau) - {z}_a^*} .
 \label{InstSolG}
\end{equation}
with
\begin{equation}
 z(\tau) = e^{-2\pi i T \tau} . \label{z-tau}
\end{equation}
For instantons ($W>0$) the complex parameters $z_a$ are all inside
the unit circle and for anti-instantons ($W<0$) they are outside.
The classical action
\begin{equation}
\mathcal{S}_d[\Phi_W] =  \frac{g}{2} |W| \label{Sclass1}
\end{equation}
is finite and independent of the complex parameters $z_a$ which
are the $2|W|$ zero modes in the problem.

In the limit where $g \rightarrow \infty$ one may generally think
in terms of a dilute gas of single instantons and anti-instantons.
One identifies $\tau_0 = \textrm{arg}\, z_1 /2\pi T$ as the {\em position}
(in time) of the single instanton whereas $\lambda = (1-|z_1|^2)\beta$ is
the {\em scale size} or the duration of the potential pulse
$i\dot{\Phi}_{\pm 1}(\tau)$. The thermodynamic potential
$\Omega_{\rm inst} = -T \ln Z$ of the dilute instanton gas~\cite{Instantons3} can be
expressed in a standard manner as an integral over $\tau_0$ and
$\lambda$ according to~\cite{PruiskenBurmistrov2,Rajaraman}
\begin{widetext}
\begin{equation} \label{Omega-inst}
 \beta \Omega_{\rm inst} = - \int_0^\beta {d\tau_0} 
 \int_0^\beta \frac{d\lambda}{\lambda^2}
 g(\lambda) D\, e^{-\frac{1}{2} g(\lambda) +
 \frac{2}{E_c (\lambda)}(T -\frac{2}{\lambda})} \cos 2\pi q.
\end{equation}
\end{widetext}
with $D= 2 e^{-\gamma -1}$ and $\gamma \approx 0.577$ the
Euler constant. Here, the quantities $g(\lambda)$ and $E_c
(\lambda)$ have the same radiative corrections as those obtained
from ordinary perturbation theory~\cite{perturb}
\begin{equation} \label{weak-RG}
 g(\lambda) = g -2 \ln \lambda \Lambda ~,~~ E_c (\lambda) = E_c
 \left( 1- \frac{2}{g} \ln \lambda \Lambda \right)
\end{equation}
with $\Lambda = g E_c / \pi^2 D$ standing for the frequency or
energy scale. From Eq. \eqref{weak-RG} we obtain the
renormalization group equations which to order $g^{-1}$ are given
by
\begin{equation}
 \beta_g = \frac{d g (\lambda)}{d\ln\lambda} = - 2 - \frac{4}{g(\lambda)} ~,~~ \beta_c
 = \frac{d\ln E_c (\lambda)}{d \ln \lambda} = - \frac{2}{g
 (\lambda)}.
\end{equation}
{Here, we have included the perturbative contribution~\cite{HZ,Beloborodov} of order $1/g(\lambda)$ into $\beta_g$.}
Based on perturbation theory alone one expects that the quantum
system is a good ``conductor" at high temperatures
\begin{equation}\label{g-weak}
 g(T) = -2 \ln \beta / \xi \gg 1
\end{equation}
and an ``insulator" at low temperatures
\begin{equation}\label{g-strong}
 g(T) = \exp\Bigl [- (\beta/\xi)^z \Bigr ] \ll 1 .
\end{equation}
Here, $z$ is a dynamical exponent that is as of yet unknown and
$\xi$ denotes the dynamically generated correlation length in the
time domain
\begin{equation} \label{xi-weak}
 \xi = \Lambda^{-1} g^{-1} e^{g/2} .
\end{equation}
These standard ideas and expectations do not reveal much about the
$\theta$ angle concept on the strong coupling side, however. For
example, there are the conflicting claims made by the
semiclassical picture of the Coulomb blockade which say that the
system displays a vanishing energy gap or a ``quantum phase
transition" when $q$ passes through half integral values, see
Section \ref{no-tunneling}. These conflicting scenarios raise
fundamental questions about the exact meaning of the topological
excitations in the problem and, in particular, the dilute gas of
instantons written in Eq. \eqref{Omega-inst}.


\section{Physical observables \label{Sec:Phys}}

As pointed out many times in our previous work, traditional
instanton results such as Eq. \eqref{Omega-inst} are of limited
significance since they merely describe the regular or
non-critical pieces of the theory which are of secondary interest.
In order to be able to understand the low energy dynamics of the
SET and, in particular, the phenomenon of charge quantization one
must develop an entirely different approach to the AES model and
reconsider the traditional renormalization group ideas in quantum
field theory all together.

Recall that conventionally one defines a renormalized theory by
specifying how the ultraviolet singularity structure of the bare
theory can be absorbed in counter terms. There are many ways of
doing this and normally, in the theory of critical exponent values
in $\epsilon$ expansions for example, one chooses a specific
scheme based on computational advantages.

The infrared problems associated with the instanton angle $\theta$
dramatically alter the physical objectives of the renormalization
group. The extensive list of studies on the AES model is in many
ways a reflection of what started many years ago in quantum field
theory.  There are the perturbative weak coupling analyzes, the
instanton investigations as well as the various different attempts
toward the strong coupling phase of the SET. Each of these
distinctly different approaches to the AES model provide different
pieces of knowledge in physics. They are completely disconnected,
however, and have physically very little in common.

The basic idea pursued in the theory of the quantum Hall effect is
to provide a unifying renormalization theory of the instanton
angle $\theta$ based on the response of the system to
infinitesimal changes in the boundary conditions. This idea is
very close to the criterion of Anderson localization originally
proposed by Thouless.~\cite{Thouless} It is also very close to 't Hooft's idea on
duality based on twisted boundary conditions~\cite{tHooft} which states that
gapless excitations must in general exist when $\theta$ passes
through odd multiples of $\pi$. Unlike these well known principles
in physics, however, one now relates the sensitivity to boundary
conditions to a set of ``physical observables" that provide a very
general definition of the renormalization behavior of the system.
In the context of the quantum Hall effect these physical
observables have previously been recognized as the macroscopic
conductance parameters of the system.

The AES model is an interesting and highly non-trivial example
where the theory of physical observables can be explored and
investigated in great detail. In this Section we show that the
problem of charge quantization in the SET is completely analogous
to the robust quantization of the Hall conductance observed in the
disordered electron gas in two dimensions. It turns out that the
AES model is extremely interesting in and of itself because of the
the long ranged nature of the tunneling term or the non-local
properties of the kernel $\gamma (\tau_{12})$ in Eq.
\eqref{SdStart-1}. For the sake of simplicity we assume throughout
the present Section that $\gamma (\tau_{12})$ is local in time and
postpone the refinements and extensions of the argument to Section
\ref{Kubo}.

As a trivial but very instructive example of our general
definition of physical observables we study the isolated
mesoscopic island in Section \ref{no-tunneling}. This naive strong
coupling example reveals much the conceptual structure of the
instanton angle $\theta$ and sets the stage for the remainder of
this investigation.

\subsection{Background fields \label{BackGround}}
Consider a fixed background matrix field $U_0 (\tau)$ or $\mathcal{Q}_0
(\tau) = U_0^{-1} \sigma_z U_0$ that varies slowly in time. We
assume that $\mathcal{Q}_0$ satisfies the classical equations of motion and
carries a small fractional topological charge, i.e. $\mathcal{Q}_0$ violates
the boundary conditions of Eq. \eqref{PBCs}. The theory in the
presence of the background field
\begin{equation} \label{Z-q-Q-0}
 Z [q; \mathcal{Q}_0] = \int_{
 \partial V} \mathcal{D} [\mathcal{Q}] e^{-\mathcal{S} [U_0 \mathcal{Q} U_0^{-1}] }
\end{equation}
then provides all the important information on the quantum system
at low energies. To relate the background field action the
appearance of an energy gap in the SET one must separate the
constant pieces in $\mathcal{Q}_0$ from the parts that couple to the the
matrix field variable $\mathcal{Q}$. Employing the split of Eq.
\eqref{split} and keeping in mind that $\mathcal{C}[\mathcal{Q}]$ is
quantized then one can write
\begin{eqnarray}
 && \exp\left\{ 2\pi i q \mathcal{C}[U_0 \mathcal{Q} U_0^{-1}] \right\} = \nonumber \\
 && ~~~~~ = \exp\left\{ 2 \pi i k(q) \mathcal{C}[\mathcal{Q}_0] + i \theta(q)
 \mathcal{C}[U_0 \mathcal{Q} U_0^{-1}] \right\} .\,\,{}
\end{eqnarray}
Using this identity one can split the theory of Eq.
\eqref{Z-q-Q-0} into pieces that are periodic and non-periodic in
the external charge $q$ according to
\begin{eqnarray} \label{Z-q-Q-0-1}
 Z [q; \mathcal{Q}_0] &=& e^{2\pi i k(q) \mathcal{C}[\mathcal{Q}_0]} ~ Z \left[\theta(q)/2\pi; \mathcal{Q}_0
 \right] .
\end{eqnarray}
It is clear that only the periodic piece probes the sensitivity of
the SET to changes in the boundary conditions. Provided the
$\gamma(\tau_{12})$ is local in time one obtaines the effective
action in $\mathcal{Q}_0$ in terms of a derivative expansion. The result is
of the same form as the AES action itself
%
\begin{eqnarray} \label{S-eff-Q-0}
 \mathcal{S}_{\rm eff} [\mathcal{Q}_0] &=& 2\pi i k(q) \mathcal{C}[\mathcal{Q}_0] + \mathcal{S}_{\theta}
 [\mathcal{Q}_0] \\ \label{S-eff-Q-1}
 \mathcal{S}_{\theta} [\mathcal{Q}_0] &=&
 \ln \frac{Z \left[\theta(q)/2\pi; \mathcal{Q}_0 \right]}{Z \left[\theta(q)/2\pi
 \right]}  \\
 &=& \int\limits_{0}^{\beta} d\tau_1  d\tau_2 ~ \gamma^\prime (\tau_{12}) \tr \mathcal{Q}_0(\tau_1) \mathcal{Q}_0(\tau_2) \notag \\
 &-&
 i \theta^\prime \mathcal{C} [\mathcal{Q}_0] + \mathcal{O}(\mathcal{Q}_0^3).\notag
\end{eqnarray}
%
except that the bare quantities $\gamma$ and $\theta(q)$ are
replaced by the effective expressions $\gamma^\prime$ and
$\theta^\prime$ respectively. As a criterion for a mass gap or
energy gap in the SET one can now state that $\mathcal{S}_{\theta}
[\mathcal{Q}_0] $ must vanish order by order in an expansion in powers of
the derivative acting on $\mathcal{Q}_0$. This means that not only the
$\gamma^\prime$ and $\theta^\prime$ are exponentially small in
$\beta$ but also the infinite series of higher order terms not
written in Eq. \eqref{S-eff-Q-1}. Under these circumstances the
effective action is given by
\begin{eqnarray} \label{S-eff-Q-0}
 \mathcal{S}_{\rm eff} [\mathcal{Q}_0] &=& {2\pi i k(q) \mathcal{C}[\mathcal{Q}_0]} .
\end{eqnarray}
In the context of the disordered electron gas one identifies this
result as the action of ``massless chiral edge excitations." The
quantity $k(q)$ is recognized as the robustly quantized Hall
conductance with sharp steps occurring at the center of the Landau
bands (i.e. $q=m+1/2$ with integer $m$).

Presently, the background field $\mathcal{Q}_0$ merely stands for a
perturbing field that takes the SET out of thermal equilibrium.
The quantity $k(q)$, however, is identified as the robustly
quantized quasi particle charge of SET. This quantity, as we shall
see in Section \ref{Kubo}, is in general very different from the
averaged charge $Q$ on the {island}.

%
\subsection{Isolated island \label{no-tunneling}}
To see these general statements at work we go back to the path
integral representation of Section \ref{Path} and consider the
simple problem with the tunneling conductance $g$ equal to zero.
The classical equation of motion of the Coulomb term of Eq.
\eqref{ScStart} is given by ${\partial^2 \Phi}/{\partial \tau^2}
=0$ which is simply solved by writing
\begin{equation}\label{sol-Phi-0}
 \Phi (\tau) =2\pi T (W+\phi ) \tau .
\end{equation}
The integer $W$ generally stands for the integral topological
sectors of the system and $-1/2 < \phi < 1/2$
denotes the perturbing background field with a fractional
topological charge. We can write
\begin{eqnarray} \label{Z-q-phi-0}
 Z [q; \phi] &=& \sum_W \exp \left\{ 2\pi i q (W+\phi) - \frac{\pi^2}{\beta E_c}
 (W+\phi)^2 \right\}
 \nonumber \\
 &=& Z[q] e^{-\mathcal{S}_{\rm eff} [\phi]} \label{Z-q-phi-1}
\end{eqnarray}
The ``effective action" $\mathcal{S}_{\rm eff} [\phi]$ in Eq.
\eqref{Z-q-phi-1} has the same general form as the original AES
theory (in the absence of tunneling)
\begin{eqnarray}\label{S-eff-phi}
 \mathcal{S}_{\rm eff} [\phi ] = - 2\pi i q^\prime \phi - \frac{\pi^2}{\beta E_c^\prime} \phi^2
 + \mathcal{O}(\phi^3)
\end{eqnarray}
except that the bare parameters $q$ and $E_c$ are now replaced by
the effective or ``observable" ones $q^\prime$ and $E_c^\prime$
respectively. It is readily seen that
\begin{eqnarray}\label{theta-prime-def}
 q^\prime &=& q +
 \frac{1}{2 \beta E_c} \frac{\partial \ln Z [q]}{\partial q},  \\
 \frac{1}{E_c^\prime} &=&
 \frac{1}{E_c} \left( 1 +
 \frac{1}{2 \beta E_c} \frac{\partial^2 \ln Z [q] }{\partial q^2} \right) .
 \label{E-c-prime-def}
\end{eqnarray}
Similar expressions can be written down for the coefficients of
the higher order terms in $\mathcal{S}_{\rm eff}$ which in general are
irrelevant.
\subsubsection{Further evaluation}
\begin{figure}[t]
\includegraphics[width=60mm]{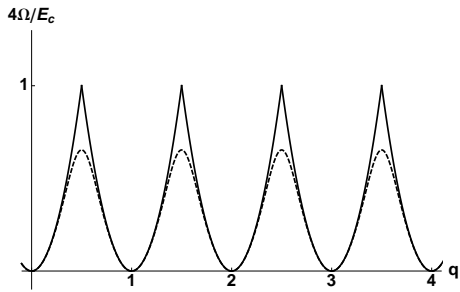}
\caption{The thermodynamic potential $\Omega$ of the isolated island at finite (dashed curve) and zero (solid curve) temperatures.} \label{Cusps}
\end{figure}

To investigate the criterion for charge quantization written in
Eq. \eqref{S-eff-Q-0} we must evaluate the observable theory of
Eqs \eqref{theta-prime-def} - \eqref{E-c-prime-def} in the limit
$T = 0$. Making use of the Poisson summation formula
\begin{equation}\label{Poisson}
 \sum_W e^{2\pi i x W} = \frac{1}{2\pi} \sum_n \delta (x-n)
\end{equation}
one can express the partition function of Eq. \eqref{Z-q-phi-0} as
a rapidly converging sum over quantum numbers $n$ according to
\begin{eqnarray} \label{Z-q-phi-2}
 Z [q; \phi]
 &=& \sum_n \exp \left\{2\pi i n \phi - \beta E_c (n-q)^2
 \right\} .
\end{eqnarray}
We immediately recognize the {grand
partition function for}
Eq.
\eqref{HamCoul} with the integer $n$ now standing for the number
of electrons on the island. The effective action can now be
written as follows
\begin{equation}\label{S-eff-phi-1}
 \mathcal{S}_{\rm eff} [\phi] = 2 \pi i \langle n \rangle \phi -
 2 \pi^2 ( \langle n^2 \rangle
 - \langle n \rangle^2 ) \phi^2 + \mathcal{O}(\phi^3) .
\end{equation}
Comparison with Eq. \eqref{S-eff-phi} shows that $q^\prime$ is
none other than the averaged charge $\langle n \rangle$ on the
{island}
and $E_c^\prime$ is related to the variance
\begin{equation}
 q^\prime = \langle n \rangle ~,~~ \frac{1}{\beta E_c^\prime} =
 8( \langle n^2 \rangle - \langle n \rangle^2 ) .
\end{equation}
To obtain explicit expressions for $q^\prime$ and $E_c^\prime$ we
follow up on Eq. \eqref{Z-q-Q-0-1} and split Eq. \eqref{Z-q-phi-2}
into periodic and non-periodic parts in $q$ according to
\begin{eqnarray} \label{Z-q-phi-3}
 Z [q; \phi] &=& e^{2\pi i k(q) \phi} ~ Z [\theta(q)/2\pi ;\phi]
 \nonumber \\
 Z \Bigl [\frac{\theta(q)}{2\pi} ;\phi\Bigr ] &=& \sum_{n^\prime}
 \exp \left\{2\pi i n^\prime \phi - \beta {E_c}
 \left( n^\prime -\frac{\theta(q)}{2\pi} \right)^2
 \right\} .\nonumber \\
 \label{Z-q-phi-4}
\end{eqnarray}
It is immediately clear that $Z [\theta(q)/2\pi;\phi]$ in
the limit $\beta \rightarrow \infty$ is independent of $\phi$. In
complete accordance with the general statement of Eq.
\eqref{S-eff-Q-0} we conclude that the {island}
for all values
of $-\pi < \theta(q) < \pi$ develops an energy gap. The quantity
$q^\prime$ or the averaged charge $\langle n \rangle$ on the
island is quantized
\begin{equation} \label{thermo-pot-1}
 q^\prime = \langle n \rangle = k(q)
\end{equation}
with sharp steps occurring at half-integral values of $q$ where
the energy gap vanishes. The thermodynamic potential
\begin{equation} \label{thermo-pot}
 \beta \Omega[q] = \beta E_c \left( \frac{\theta (q)}{2\pi} \right)^2
\end{equation}
displays a ``cusp" at half-integral values of $q$ indicating that
the transition is a first order one (see Fig.~\ref{Cusps}).

%
\subsubsection{Renormalization}
\begin{figure}[t]
\includegraphics[width=60mm]{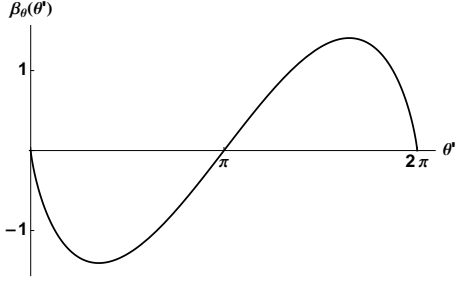}
\caption{$\beta_\theta$ function (Eq.~\eqref{strong-RG}) for the isolated island.} \label{FigBeta0}
\end{figure}

This takes us to the most important part of this exercise which is
to show that the physical observables generally define the
renormalization behavior of the {island}
at finite $T$. Notice
that Eq. \eqref{Z-q-phi-4} is dominated by the terms with
$n^\prime =0,\pm 1$. Write
\begin{eqnarray}
 Z [\theta(q)/2\pi; \phi] &=& Z [\theta(q)/2\pi]
 e^{-\mathcal{S}_{\rm eff}^0 [\phi]} \\\label{S-eff-isol}
 \mathcal{S}_{\rm eff}^0 [\phi] &=& -i \theta^\prime \phi -
 \frac{\pi^2}{\beta E_c^\prime} \phi^2 + \mathcal{O}(\phi^3)
\end{eqnarray}
then the explicit results for the thermodynamic potential $\Omega
[q]$ and the physical observables $\theta^\prime$ and $E_c^\prime$
can be written as follows
\begin{eqnarray}
 \beta \Omega [q] &=& - \ln Z [{\theta(q)}/{2\pi}] \nonumber \\
 &=& \frac{1}{4}(\beta {E_c}) \left( 1-{\Delta_0}\right)^2
 -\ln \left[ 1+e^{-(\beta E_c) \Delta_0} \right]
 \label{Omega-q-00} \\
 \theta^\prime &=& \theta(q) - \frac{\pi}{E_c} \frac{\partial \Omega[q]}{\partial
 q}\nonumber\\
 &=& \pm \frac{2\pi}{e^{(\beta E_c) \Delta_0}+1} \label{theta-prime-00} \\
 \frac{1}{\beta E^\prime_c} &=& \frac{1}{\beta E_c} \left( 1 - \frac{1}{2E_c}
 \frac{\partial^2 \Omega [q]}{\partial q^2} \right) \nonumber\\
 &=& \Bigl |\frac{\theta^\prime}{2\pi}\Bigr |
 \left( 1-\Bigl |\frac{\theta^\prime}{2\pi}\Bigr | \right) .
 \label{E-c-prime-00}
\end{eqnarray}
Here, $\pm$ denotes the sign of $\theta(q)$ and $\Delta_0$ is
recognized as the the dimensionless energy gap of the {island}
which vanishes near the critical point according to
\begin{equation}
 \Delta_0 = \left( 1 - \left| \frac{\theta(q)}{\pi} \right|
 \right).
\end{equation}
Finally, we express the response quantity $\theta^\prime$ in
differential form and obtain (see Fig.~\ref{FigBeta0})
\begin{equation}\label{strong-RG}
 \beta_\theta (\theta^\prime) = \frac{d \theta^\prime}{d\ln \beta} =
 \frac{\theta^\prime}{2\pi} \left[ 2\pi -|\theta^\prime| \right]
 \ln \left[ \frac{|\theta^\prime|}{2\pi - |\theta^\prime|}\right] .
\end{equation}
This result clearly translates the physics of the isolated island in the language of the renormalization group. Notice that the quantity $E_c^\prime$ in Eq. \eqref{E-c-prime-00} does not lead to more complex renormalization behavior since it is expressed in terms of $\theta^\prime$ alone.
The same is true for the higher terms in Eq. \eqref{S-eff-isol}.

We identify two
different kinds of strong coupling fixed points, a stable one at
$\theta^\prime=0$ and a critical one at $\theta^\prime=\pm \pi$.
\begin{enumerate}
\item
Near the critical fixed point $\theta^\prime=\pm\pi$ we find
\begin{equation}\label{pi-theta-prime}
 \beta_\theta (\theta^\prime)
 = \pm \pi +\theta^\prime
\end{equation}
which is a standard result for a first order transition in one
dimension. Eq. \eqref{pi-theta-prime} determines the energy gap
exponent $\nu$ according to
\begin{equation}
 \frac{1}{\nu} =
 \left[ \frac{\partial \beta_\theta}{\partial \theta^\prime}
 \right]_{\theta^\prime = \pm \pi}=1 .
\end{equation}
Eqs \eqref{S-eff-phi-1} and \eqref{E-c-prime-00} tell us that near criticality the charge on the island is broadly distributed, i.e. the fluctuations are of the same order of magnitude as the averaged value $\theta^\prime$.
\item
Near the stable fixed point at $\theta^\prime =0$ we find
\begin{equation}\label{stable-strong}
 \beta_\theta (\theta^\prime) = \theta^\prime \ln |\theta^\prime|
\end{equation}
indicating that the averaged charge $q^\prime$ on the {isolated island}
is robustly quantized with corrections that are exponentially
small in $\beta$, i.e.
\begin{equation}
 q^\prime = k(q)+\frac{\theta^\prime}{2\pi} = k(q) \pm
 e^{-\beta E_c \Delta_0} .
\end{equation}
Similarly, the root-mean-square fluctuations in $\theta^\prime$ as well as the higher order moments all render exponentially small in $\beta$.
\end{enumerate}
%

\section{General response theory \label{Kubo}}
Armed with the insights obtained from the isolated island we next address the AES theory with finite values of $g$. To discuss the tunneling term $\mathcal{S}_d [\Phi]$ with varying boundary conditions on the $\Phi$ field one must generalize the
expression for the kernel $\alpha(\tau_{12})$ in Eq. \eqref{alpha-tau} which is
periodic in time. Write
\begin{eqnarray}\label{alpha-1}
 {\alpha}_\phi (\tau_{12}) = \frac{T}{\pi} \sum_n e^{-i\left( \omega_n + 2\pi T
 \phi\right) \tau_{12}} ~|\omega_n +2\pi T \phi|
\end{eqnarray}
with $-1/2 < \phi < 1/2$. The appropriate result
for the tunneling term $\mathcal{S}_d $ in Eq. \eqref{alpha-tau} is then obtained if one replaces
$\alpha(\tau_{12})$ by the following expression
%
\begin{eqnarray}
 {\alpha} (\tau_{12}) & \rightarrow & e^{i\left(2\pi T \phi\right) \tau_{1}}
 {\alpha}_\phi (\tau_{12}) e^{-i\left(2\pi T \phi\right) \tau_{2}} \label{alpha-2}\\
&=&{\alpha} (\tau_{12}) + 2 T^2 |\phi| - 2 i T^2 \phi \cot (\pi T \tau_{12}) .\notag
\end{eqnarray}
%
Eq. \eqref{alpha-2} essentially tells us that one cannot insert a
background field $\mathcal{S}_d [\Phi]
\rightarrow \mathcal{S}_d [\Phi+\Phi_0]$ with $\Phi_0 = 2\pi T
\phi$ carrying a fractional topological charge unless one changes
the kernel $\alpha(\tau_{12})$ into $\alpha_\phi (\tau_{12})$. Given
Eq. \eqref{alpha-2} it is straightforward to discuss the effect of
the more general background field
\begin{equation}\label{Phi-0}
 \Phi_0 (\tau) = (\omega_m + 2\pi T \phi ) \tau
\end{equation}
and the result can be written as follows
%
\begin{gather}
 \mathcal{S}_d [\Phi + \Phi_0 ] = \frac{g}{4} \int\limits_0^\beta d\tau_{1}d\tau_2\,
 e^{i\Phi(\tau_1) -i \Phi(\tau_2) + i\Phi_0 (\tau_1) -i \Phi_0 (\tau_2)}\notag \\
 \times\alpha_\phi (\tau_{12}) = \frac{g}{4\pi} \int\limits_0^\beta d\tau_{1}d\tau_2\, e^{i\Phi(\tau_1) -i \Phi(\tau_2)}\notag
\\ \times  T \sum_n e^{-i\omega_n \tau_{12}} ~|\omega_n + \dot{\Phi}_0| .
 \label{S-d-Phi-Phi-1}
\end{gather}
%
Notice that Eq. \eqref{Phi-0} now satisfies
the classical equations of motion of the AES theory as a whole,
i.e. not for only the isolated {island} as discussed in the previous Section but
also for the theory in the presence of tunneling. We will next embark
on the distinctly different ways of handling the background field
methodology depending on the topological charge of the field
$\Phi_0$.
\subsubsection{$\Phi_0$ with fractional topological charge \label{response-1}}
By taking $\Phi_0 = 2\pi T \phi \tau$ or $\omega_m = 0$ then Eq.
\eqref{Phi-0} can directly be used to probe the sensitivity of the
SET to changes in the boundary conditions. Introducing the
two-point correlation function
\begin{gather}
 D(i\omega_n) = T \int_0^\beta \int_0^\beta {d\tau_1 d\tau_2}
 e^{i\omega_n\tau_{12}}\left \langle e^{-i
 \Phi(\tau_1)+i\Phi(\tau_2)}\right \rangle .\label{DMdef}
\end{gather}
then to lowest orders in the $\phi$ we obtain the total effective
action
\begin{eqnarray}
 \mathcal{S}_{\rm tot} [\phi] &=& \mathcal{S}_{\rm eff} [\phi] +
 \nonumber \\
 &+& \frac{g}{4\pi} \sum_n D (i\omega_n) \left( |\omega_n + 2\pi T
 \phi | - | \omega_n | \right) \label{S-tot}
\end{eqnarray}
with $\mathcal{S}_{\rm eff} [\phi]$ given by Eq. \eqref{S-eff-phi-1} and
below. Keeping in mind that $-1/2 < \phi < 1/2$ we split the sum in
Eq. \eqref{S-tot} in $n=0$ and $n \ne 0$ parts and we immediately
obtain
\begin{eqnarray}\label{S-eff-phi}
 \mathcal{S}_{\rm tot} [\phi ] = \frac{g^\prime}{2} |\phi|
 - 2\pi i q^\prime \phi
 + \delta \mathcal{S}_{\rm tot} [\phi ]
\end{eqnarray}
where $\delta \mathcal{S}_{\rm tot}$ stands for all the higher order
terms in $\phi$. The physical observables $g^\prime$ and
$q^\prime$ are given as follows
\begin{eqnarray}\label{g-prime-def}
 g^\prime &=& g T D(i0) ,\\
 q^\prime &=& Q
 - \frac{g}{2\pi} T \sum\limits_{n>0} \Im D (i\omega_n).
 \label{theta-prime-def}
\end{eqnarray}
Here we have introduced the quantity
\begin{equation}
Q= q + \frac{i\langle \dot{\Phi} \rangle }{2E_c} = q -
 \frac{1}{2E_c} \frac{\partial \Omega[q]}{\partial q}  \label{Qdef}
\end{equation}
which generally stands for the averaged charge on the island. We
see that in the presence of tunneling the averaged charge
$Q$ is different from $q^\prime$ which we now identify
with the quasi particle charge of the SET. We expect that the new
quantity $q^\prime$ is quantized and, along with that, the
quantity $g^\prime$ as well as all the higher dimensional terms in
$\delta \mathcal{S}_{\rm tot}$ render exponentially small in the limit
where $\beta$ goes to infinity.
\subsubsection{Higher dimensional terms}
To obtain the leading order corrections in $\delta
\mathcal{S}_{\rm tot}$ one needs the four point correlation function
\begin{eqnarray}
 D(i\omega_n , i\omega_m) &=& T^2 \int_{12} \int_{34}
 e^{i\omega_n\tau_{12}+i\omega_n\tau_{34}} \label{DMdef-2} \\
 &\times& \left \langle e^{-i \Phi(\tau_1)+i\Phi(\tau_2)} ~
 e^{-i\Phi(\tau_3)+i\Phi(\tau_4)}\right \rangle_{cum} .\notag
\end{eqnarray}
The effective action $\delta \mathcal{S}_{\rm tot}$ up to the third
order in $\phi$ can be written as follows
\begin{eqnarray}\label{delta-S-eff-phi}
 \delta \mathcal{S}_{\rm tot} [\phi ] = \frac{\pi^2}{\beta E_c^\prime}
 \phi^2 + \frac{2\pi i}{\beta F^\prime_c} \phi |\phi| + \mathcal{O}(\phi^3)
\end{eqnarray}
where
\begin{eqnarray}\label{E-c-prime}
 \frac{1}{\beta E^\prime_c} &=& ~\frac{1}{\beta E_c} \frac{\partial
 }{\partial q}\Bigl (2q^\prime-Q\Bigr ) + \frac{g^2 T^2}{8\pi^2} D(i0,i0)
\\
&-& \frac{g^2 T^2}{8\pi^2} \sum\limits_{n,m\neq 0} \sign(\omega_n \omega_m) D(i\omega_n,i\omega_m) ,
 \notag \\
 \frac{1}{\beta F^\prime_c} &=& \frac{1}{4\beta E_c} \frac{\partial
 g^\prime}{\partial q} + \frac{ig^2T^2}{8\pi} \sum\limits_{n\neq 0} \sign\omega_n D(i\omega_n,i0) \label{G-1-2}.
\end{eqnarray}
%
%

A more detailed discussion of the higher order terms will be presented elsewhere.~\cite{Elsewhere}

\subsubsection{$\Phi_0$ with integral topological charge \label{response-2}}
A less obvious way of probing the energy gap in the SET is
obtained by putting $\phi=0$ and, instead, we consider background
fields with an integral topological charge only, i.e. $\Phi_0 =
\omega_m \tau$. Notice that this choice of $\Phi_0$ is a special
case of the instanton solution of Eq. \eqref{InstSolG} with $W=m$
but with all the parameters $z_\alpha$ put equal to zero. Even
though this background field $\Phi_0$ can formally be absorbed in
a redefinition of the $\Phi$ field one can nevertheless proceed
and define the effective action $\mathcal{S}_{\rm tot} [\Phi_0]$ by
expanding in powers of $\Phi_0$ or $\omega_m$. Provided one finds
a way to analytically continue the discrete Matsubara frequencies
to fractional or infinitesimal values the final results are again
a measure for the sensitivity of the SET to changes in the
boundary conditions.

To start we consider the effective action at a tree level
\begin{equation}\label{class-action}
 \mathcal{S} [\Phi_0] = \frac{g}{2} |m| -
 2\pi i q m + \frac{\pi^2}{\beta E_c} m^2 .
\end{equation}
We expect that the exact result retains the general form of Eq.
\eqref{class-action} except that the bare parameters $g$, $q$ and
$E_c$ are replaced by effective or observable ones. To lowest
orders in $m$ one can write
\begin{eqnarray}
 \mathcal{S}_{\rm tot} [m] &=& \mathcal{S}_{\rm eff} [m] +
 K(i\omega_m) - K(i0) \label{S-tot-1}
\end{eqnarray}
where $\mathcal{S}_{\rm eff} [m]$ is the same as Eq. \eqref{S-eff-phi-1}
with $\phi$ replaced by $m$. We have introduced the quantity
%
\begin{equation}
 K(i\omega_n) = -\frac{g}{4\beta} \int\limits_0^\beta {d\tau_1
 d\tau_2} ~e^{i\omega_n\tau_{12}}
 ~\alpha(\tau_{12})~
 \left\langle e^{i\Phi(\tau_1) - i\Phi(\tau_2)} \right\rangle .
 \label{KMdef}
\end{equation}
%
To expand this theory in terms of a series in powers of $\omega_m$
we make use of the analytic properties of response functions.
Specifically, following the standard prescription $i \omega_m
\rightarrow \omega + i0^+$ we analytically continue the discrete
set of imaginary frequencies $i\omega_m$ in Eq. \eqref{KMdef} to
real ones $\omega$ and subsequently we can take the limit $\omega
\rightarrow 0$, see Section \ref{More}. The following total result
is obtained for the effective action up to order $m^3$
\begin{eqnarray}\label{S-eff-m}
 \mathcal{S}_{\rm tot} [ m ] = -\frac{g^\prime}{2} |m|
 - 2\pi i q^\prime m + \frac{\pi^2}{\beta E_c^\prime} m^2
 + \frac{2 \pi i}{\beta F_c^\prime} m |m|.\nonumber \\
\end{eqnarray}
Here, the quantities $q^\prime$ and $g^\prime$ are given in terms
of Kubo-like expressions as follows
\begin{eqnarray}
 g^\prime &=& 4\pi \Im \frac{\partial K^R(\omega)}{\partial\omega} \label{gPrBack}\\
 q^\prime &=& Q
 + \Re \frac{\partial K^R(\omega)}{\partial\omega} .\label{thetaPrBack}
\end{eqnarray}
$K^R (\omega)$ denotes the analytic continuation of $K(i\omega_m)$
and the limit $\omega \rightarrow 0$ is understood. As before the
$Q$ denotes the averaged charge on the island {(see Eq.~\eqref{Qdef}).}
%
%
Similar results can be written down for $E_c^\prime$ as well as
$F_c^\prime$ but the expressions are considerably more complex
than those of Eqs \eqref{E-c-prime} - \eqref{G-1-2} and the
details will be presented elsewhere.

Even though the physical observables of this Section are formally
different from those in the preceding Section they should nevertheless 
define the same renormalization behavior of the SET. In particular, in the
presence of an energy gap the physical observables in both Eq. \eqref{S-eff-phi}
and Eq. \eqref{S-eff-m} should all scale to zero as $\beta$ goes to
infinity except for the quantity $q^\prime$ that can take on
arbitrary integral values. The main advantage of Eqs \eqref{S-eff-phi} - 
\eqref{thetaPrBack}, however, is that they directly lay the bridge between
the background field methodology on the one hand, and results
obtained from ordinary linear response theory on the other, see
Section \ref{PhysObs}.
%
%
\subsection{More about response functions \label{More}}

The function $K^R(\omega)$ can elegantly be expressed in terms of
the retarded propagator $D^R(\omega)$ which is the analytic
continuation of $D(i\omega_n)$ in Eq. \eqref{DMdef}. In
Appendix~\ref{App:AnalCont} we derive the following relation
%
\begin{gather}
 K^R(\omega) =g \int\limits_{-\infty}^\infty \frac{d \epsilon_1
 d \epsilon_2}{4\pi^3} ~ \epsilon_2
 \frac{n_b (\epsilon_2)-n_b (\epsilon_1)}{\epsilon_1 -\epsilon_2 +\omega+i
 0^+} \Im D^R(\epsilon_1)\label{KRdef}
\end{gather}
%
where $n_b(\epsilon) = [\exp(\beta \epsilon)-1]^{-1}$ denotes the
Bose-Einstein distribution. A detailed computation of $K^R(\omega)$ in the weak and strong coupling regimes is presented in Appendix~\ref{genKR} . 

Given the function $K^R(\omega)$ one can obtain the response parameters $g^\prime$ and $q^\prime$ from Eqs.~\eqref{gPrBack} and \eqref{thetaPrBack}. However,  
it is more convenient to express these quantities directly in terms of $D^R(\omega)$
according to
\begin{eqnarray}
 g^\prime &=& g\int\limits_{-\infty}^\infty \frac{d \epsilon}{\pi} ~ \epsilon
 \frac{\partial
 n_b }{\partial \epsilon} ~ \Im D^R(\epsilon) , \label{gPrKor}\\
%
 q^\prime &=& Q + g \int\limits_{-\infty}^\infty \frac {d \epsilon}{4\pi^2} ~
 \frac{\partial \epsilon n_b}{\partial \epsilon} ~ \Re
 D^R(\epsilon).
 \label{qPrKor}
\end{eqnarray}
\subsection{Linear response \label{PhysObs}}
In this Section we establish the contact between the response
quantities $g^\prime$ and $q^\prime$ and the well known
expressions for the SET conductance $G$ and the non-symmetrized
current noise $S_I$.
\subsubsection{The SET conductance}
If one applies a voltage difference $V=V_r-V_l$ between the
reservoirs then the tunneling part $\mathcal{H}_T^{(l,r)}$ of
Eq.~\eqref{HamStart} becomes time dependent~\cite{Mahan}
\begin{gather}
 \mathcal{H}_T^{(s)} = X^{(s)} e^{-i e V_s t} +
 X^{(s)\dag} e^{i e V_s t}, \notag \\ X_s =
 \sum_{k\alpha} t_{k\alpha}^{(s)} a_{k}^{(s)\dag} d_{\alpha} .
\end{gather}
The operator for the current $I_s$ that flows from a reservoir to the island can be expressed as follows
\begin{gather}
 \mathcal{I}_s = e \frac{d}{dt} \sum_{k} a^{(s)\dag}_{k}
 a_{k}^{(s)}= -i e X^{(s)} e^{-i e V_s t} + h.c.
\end{gather}
To the lowest order in $1/N_{ch}^{(s)}$ we find
\begin{eqnarray}
 I_s &=& - i \int_{-\infty}^{t} d t^\prime\langle
 [\mathcal{I}^{(\alpha)}(t),
 \mathcal{H}_T^{(s)}(t^\prime)]\rangle \nonumber \\
 &=& -2 e \Im K_s^R(-eV_s) .\label{CurOperII}
\end{eqnarray}
The retarded correlation function is given by
\begin{equation}
 K_s^{R}(\omega) = i \int_{0}^\infty dt\, e^{i\omega t}\langle
 [X^{(s)}(t),X^{(s)\dag}(0)]\rangle
\end{equation}
and the corresponding Matsubara correlation function by
\begin{equation}
 K_s(i\omega_n) = \int_0^\beta d \tau e^{i\omega_n \tau}
 \langle T_\tau X^{(s)}(\tau) X^{(s)\dag}(0)\rangle .
\end{equation}
Repeating the same steps that led to the AES action starting from
the Hamiltonian of Eq. ~\eqref{HamStart} we obtain
\begin{gather}
 K_s(i\omega_n) = -\frac{g_s}{4\beta}\int_0^\beta
 {d\tau_1
 d\tau_2} e^{i\omega_n\tau_{12}} \alpha(\tau_{12})
 D(\tau_{21}).\label{KRaM0}
\end{gather}
Comparison with Eq.~\eqref{KMdef} yields $K_s(i\omega_n) =
(g_s/g) K(i\omega_n)$ or, equivalently,
\begin{equation}
 K^R_s(\omega) = (g_s/g)K^R(\omega).\label{KaK}
\end{equation}
Based on the continuity equations for the current $I = I_l = -I_r
= G V$ we finally find the SET conductance $G$ in units of
$[e^2/h]$ according to~\cite{Cur,Schon,GG2}
\begin{gather} \label{SET-G}
 G =  \frac{g_l g_r}{(g_l+g_r)^2} g^\prime
\end{gather}
with $g^\prime$ given by Eq. \eqref{gPrBack} or \eqref{gPrKor}.
Therefore, except for the constant $g_lg_r/(g_l+g_r)^2$ the
conductance $G$ is none other than the observable $g^\prime$ that
measures the sensitivity of the SET to changes in the boundary
conditions.
\subsubsection{Quantum current noise}
Similar to Eq.~\eqref{CurOperII} we obtain the real part of the
retarded correlation function as follows
\begin{gather}
 \Re K^R_s(-eV_\alpha)=\frac{i}{2e^2}  \int_{-\infty}^t d
 t^\prime \langle [\mathcal{I}_s(t),
 \mathcal{I}_s(t^\prime)] \rangle .\label{deltaI}
\end{gather}
The quantity $q^\prime$ in Eq. \eqref{thetaPrBack} or
\eqref{qPrKor} can therefore be expressed in terms of the
current-current correlation function~\cite{Letter}
\begin{gather}
 q^\prime = Q - i \frac{(g_l+g_r)^2}{2g_l g_r}
 \frac{\partial}{\partial V} \int_{-\infty}^0 dt \langle [
 \mathcal{I}(0), \mathcal{I}(t)]\rangle
 \label{qPrNoise}
\end{gather}
in the limit where $V$ goes to zero. We have thus found a novel
interpretation of the so-called antisymmetric current-current
correlation function that in different physical contexts has
attracted a considerable amount of interest over the
years~\cite{Blanter}. Introducing the non-symmetrized current
noise~\cite{Imry}
\begin{equation}
 S_I(\omega,V) = \int_{-\infty}^\infty dt \, e^{-i\omega t} \langle
 \mathcal{I}(t) \mathcal{I}(0) \rangle
\end{equation}
then one can also write
\begin{gather}
 q^\prime = Q + \frac{(g_l+g_r)^2}{g_l g_r}
 PV \int
 \frac{d\omega}{2\pi} ~\frac{1}{\omega}~ \frac{\partial S_I(\omega,V)}{\partial
 V}. \label{qSI}
\end{gather}
Here, $PV$ denotes the principal value and the limit $V
\rightarrow 0$ is understood. Eqs \eqref{SET-G} and \eqref{qSI}
are amongst the most significant results of this investigation.


\section{Weak coupling regime, $g^\prime \gg 1$ \label{Sec:Weak}}
\subsection{Perturbation theory}
At a gaussian level the AES action in frequency representation is
given by
\begin{equation}
 \mathcal{S}_{0} = g \sum_{n>0}\left ( n + \frac{2\pi^2 T}{g E_c}
 n^2 \right ) \Phi_n \Phi_{-n} .\label{Squart}
\end{equation}
To lowest order in an expansion in $g$ the following result for
$D(i\omega_n)$ is obtained
\begin{gather}
 D(i\omega_n) = \beta \left [1-
 \frac{2}{g}\sum_{s>0}\frac{1}{s+2\pi^2 T s^2/(gE_c)} \right ]
 \delta_{n,0} \notag
 \\ + \frac{2\pi i}{g} (1-\delta_{n,0})\left (\frac{1}{i|\omega_n|} -
 \frac{1}{i|\omega_n|+ig E_c/\pi}\right) .
\end{gather}
Using the representation $\delta_{n,0} = \lim\limits_{\eta\to 0}
\eta (i\omega_n+\eta)^{-1}$ one can perform analytic continuation
to real frequencies and the retarded correlation function becomes
\begin{gather}
 D^R(\omega) = \beta \left [1- \frac{2}{g}\ln
 \frac{gE_ce^\gamma}{2\pi^2 T} \right ] \lim\limits_{\eta\to 0}
 \frac{\eta}{\omega+\eta+i0^+} \notag\\ + \frac{2\pi i}{g} \left
 (\frac{1}{\omega+i0^+} - \frac{1}{\omega+ig E_c/\pi}\right)
 .\label{DRPert}
\end{gather}
Having carried out integration in Eqs.~\eqref{gPrKor} and
\eqref{qPrKor} with the help of the identity:
\begin{equation}
\int_0^\infty \frac{d x}{x^2+\pi^2 z^2}\frac{x^2}{\sinh^2 x} =\frac{1}{2|z|}-1+ |z|
\psi^\prime(1+|z|),
\end{equation}
where $\psi(z)$ denotes the Euler di-gamma function, we obtain
\begin{gather}
g^\prime(T) = g  - 2 \ln \frac{g E_c e^{\gamma+1}}{2\pi^2 T},\quad
q^\prime(T) = q .\label{Pert}
\end{gather}
Here, $\gamma=-\psi(1)\approx 0.577$ denotes the Euler constant.
The result for $g^\prime$ was originally obtained in Ref.~[\onlinecite{perturb}]
more than two decades ago. The quantity
$q^\prime$, on the other hand, is unaffected by the quantum fluctuations to
any order in an expansion in powers of $1/g$. To
establish the renormalization of $q^\prime$ ($\theta$-renormalization) it is
necessary to include the non-perturbative effects of instantons.

\subsection{Instantons}
Since the infrared of the dilute instanton gas is well defined 
one can proceed and evaluate the integrals in Eq.~\eqref{Omega-inst}. 
This leads to the much simpler expression~\cite{Instantons3,Beloborodov}
\begin{equation}
 \beta \Omega_\textrm{inst} = - \frac{g^2 }{\pi^2} \beta E_c e^{-g/2} \ln
 \frac{\beta E_c}{2\pi^2 e^\gamma } \cos 2\pi q .\label{ZresInst}
\end{equation}
With the help of Eq.~\eqref{Qdef} we immediately find the temperature
dependence of the average charge on the island and the result is 
\begin{equation}
 Q(T) = q - \frac{g^2}{\pi} e^{-g/2} \ln \frac{E_c}{2\pi^2 e^\gamma
 T} \sin 2\pi q . \label{QresInst}
\end{equation}
To find the quantities $q^\prime$ and $g^\prime$,
however, we still have to evaluate the instanton contribution to the correlation function $D(i\omega_n)$. For this purpose we first consider the expectation of an 
arbitrary operator $\mathcal{O}$ which can be expanded to lowest order in the topological sectors $W=\pm 1$ according to
\begin{equation}
 \langle \mathcal{O}\rangle \approx \frac{1}{Z[q]} \left(
 \mathcal{O}_0 + e^{2\pi i q} \mathcal{O}_1 + e^{-2\pi i q} \mathcal{O}_{-1} \right) \label{Oser}
\end{equation}
where
\begin{equation}
 \quad \mathcal{O}_W = \int_{
 \Phi (\beta) = \Phi (0) + 2\pi W }
 \mathcal{D} \Phi(\tau)~\mathcal{O}(\Phi)~e^{-\mathcal{S}_d [\Phi]
 - \mathcal{S}_c [\Phi] } .
\label{Oser-1} 
\end{equation}
Similarly, we expand the partition function according to
\begin{equation}
 Z[q] \approx Z_0 \left(
 1 + e^{2\pi i q} \frac{Z_1}{Z_0} + e^{-2\pi i q} \frac{Z_{-1}}{Z_0} \right) \label{Oser-2}
\end{equation}
Eq. \eqref{Oser} can therefore be split in a $W=0$ part and an instanton part
\begin{gather}
 \langle \mathcal{O}\rangle \approx \langle \mathcal{O}\rangle_0 +
 \langle \mathcal{O} \rangle_\textrm{inst} .
\end{gather}
Here, $\langle \mathcal{O}\rangle_0 = \mathcal{O}_0/Z_0$ and
\begin{gather}
 \langle \mathcal{O} \rangle_\textrm{inst} = e^{2\pi i q} \frac{\mathcal{O}_{1}
 - \langle \mathcal {O}\rangle_0 Z_1}{Z_0} +  e^{-2\pi i q}
 \frac{\mathcal{O}_{-1} - \langle \mathcal{O}\rangle_0 Z_{-1}}{Z_0} .\label{Oser-3}
\end{gather}
In the semi classical evaluation of Eq. \eqref{Oser-3} it suffices to replace 
the operator $\mathcal{O} [\Phi]$ in the integrand of Eq. \eqref{Oser-1} by its classical value $\mathcal{O} [\Phi_{W}]$. The result for Eq. \eqref{Oser-3} can 
then be written in the typical instanton form
\begin{widetext}
\begin{equation} \label{Omega-inst-1}
 \langle \mathcal{O} \rangle_\textrm{inst} = \sum_{W=\pm 1} \int_0^\beta {d\tau_0} 
 \int_0^\beta \frac{d\lambda}{\lambda^2} ~\bigl [ \mathcal{O} [\Phi_W] 
 -\langle  \mathcal {O}\rangle_0 \bigr ]
 ~g(\lambda) D ~\exp\left  \{-\frac{1}{2} g(\lambda) +
 \frac{2}{\beta E_c (\lambda)}\left (1 -\frac{2\beta }{\lambda}\right ) + 2\pi i q W \right \} 
\end{equation}
\end{widetext}
where $\mathcal{O} [\Phi_{\pm 1}]$ generally depends on the position $\tau_0$
and scale size $\lambda$ of the instanton/anti-instanton.

We next apply these general results to the correlation function $D(i\omega_n)$.
The operator specific parts of Eq. \eqref{Omega-inst} are computed to be, to the leading order in $1/g$,
\begin{gather}
 \int_0^\beta {d\tau_0} ~\bigl [ \mathcal{O} [\Phi_W] 
 -\langle  \mathcal {O}\rangle_0 \bigr ]  \hspace{3.5cm} \, {}\\
 =\beta \left\{ 
 - \left( \frac{\lambda}{\beta} \right) \delta_{n,0} 
 + \left(1-\frac{\lambda}{\beta} \right)^{(|n|-1)}
 \left( \frac{\lambda}{\beta} \right)^2 \Theta(n W)\right\} \notag
\end{gather}
with $\Theta$ denoting the Heaviside step function.
Inserting this result in Eq. \eqref{Omega-inst} and performing the integral
over $\lambda$ we find the following result for the instanton part
\begin{gather}
 D_\textrm{inst}(i\omega_n) = -\frac{g^2 E_c}{\pi^2 T^2} e^{-g/2}
 \Biggl [ \delta_{n,0}\cos2\pi q -  \pi i T e^{ 2\pi i q \sign n}
 \notag
 \\ \,{}\hspace{1.5cm} \times  (1-\delta_{n,0})\left (\frac{1}{i|\omega_n|} -
 \frac{1}{i|\omega_n|+ 2\pi i T}\right ) \Biggr ] .\label{DRInst}
\end{gather}
Performing analytic continuation to real frequencies we obtain
\begin{gather}
 D^R_\textrm{inst}(\omega) = -\frac{g^2 E_c}{\pi^2 T^2} e^{-g/2}
 \Biggl [ \cos2\pi q \lim\limits_{\eta\to 0}
 \frac{\eta}{\omega+\eta+i0^+}  \notag
 \\ \,{}\hspace{1cm} -  \pi i T e^{i 2\pi q} \left (\frac{1}{\omega+i0^+} -
 \frac{1}{\omega+i 2\pi T}\right ) \Biggr ] .
\end{gather}
Using Eqs.~\eqref{gPrKor} and \eqref{qPrKor} we find the
following non-perturbative corrections to $g^\prime$ and $q^\prime$
\begin{gather}
 g^\prime_\textrm{inst} = - \frac{g^3 E_c}{6 T} e^{-g/2}\cos 2\pi
 q , \label{gPrInst}
 \\
 q^\prime_\textrm{inst} = Q(T) - \frac{g^3 E_c}{24\pi T}
 e^{-g/2}\sin 2\pi q . \label{qPrInst}
\end{gather}
Here, the expression for $Q(T)$ is given by Eq.~\eqref{QresInst}.

Combining the perturbative and non-perturbative contributions of 
Eqs.~\eqref{Pert}, \eqref{QresInst}, \eqref{gPrInst} and \eqref{qPrInst} we 
obtain the final total result for the temperature dependence of $g^\prime$ and $q^\prime$
\begin{gather}
 g^\prime(T) = g  - 2 \ln \frac{g E_c e^{\gamma+1}}{2\pi^2 T}
 -\frac{g^3 E_c}{6T} e^{-g/2} \cos 2\pi q ,\label{gPrT}\\
 q^\prime(T) = q - \frac{g^3 E_c}{24\pi T} \left [ 1 + \frac{24
 T}{g E_c} \ln \frac{E_c}{2\pi^2 e^\gamma T} \right ] e^{-g/2} \sin
 2\pi q . \label{qPrT}
\end{gather}
Several remarks are in order. First of all, we notice that 
the amplitude of the the oscillations in $q^\prime$ with varying $q$
are much larger than those in the averaged charge $ Q(T)$.
Eqs ~\eqref{gPrT} and \eqref{qPrT} are generally valid in the weak 
coupling phase of the SET $g \ll 1$ such that
$T \gg g^3 E_c e^{-g/2}$. The results are completely analogous to the instanton corrections to the conductances $\sigma_{xx}^\prime$ and $\sigma_{xy}^\prime$
in the theory of the quantum Hall effect~\cite{PruiskenNucl,Comment} that have recently been investigated experimentally.~\cite{Murzin} It should be mentioned that Eq.~\eqref{gPrInst} coincides with the earlier computations reported in Ref.~[\onlinecite{Cond}].
\subsection{$\theta$ renormalization \label{Sec:RGweak}}
To leading order in $1/g$ one can express Eqs.~\eqref{gPrT} and
\eqref{qPrT} in the following manner
\begin{gather}
 g^\prime(T) = g(T) - D g^2(T)  e^{-g(T)/2} \cos 2\pi q ,
 \label{gPrT1}\\
 q^\prime(T) = q - \frac{D}{4\pi} g^2(T)  e^{-g(T)/2} \sin 2\pi q .
 \label{qPrT1}
\end{gather}
Here, $D = (\pi^2/3) e^{-\gamma-1}\approx 0.68$ is a numerical constant and
\begin{equation}
 g(T)= g  - 2 \ln \frac{g E_c}{6 D T}\label{gTres}
\end{equation}
contains the perturbative quantum corrections to leading order in
$1/g$. It is important to emphasize that same results of Eqs.~\eqref{gPrT1} 
and \eqref{qPrT1} are obtained if one employs the much simpler expressions 
for $g^\prime$ and $q^\prime$ defined in Eqs ~\eqref{g-prime-def} and \eqref{theta-prime-def}. The only difference is the numerical value of $D$ which now equals $D=2\exp(-\gamma)$. At the same time, the charging energy $E_c$ in 
Eq.~\eqref{gTres} is replaced by $(6/\pi^2)E_c$.

Expressing Eqs.~\eqref{gPrT1} and \eqref{qPrT1} in differential form 
\begin{eqnarray}
 \beta_g &=& \frac{d g^\prime}{d \ln \beta} = - 2 -
 \frac{4}{g^\prime} -D g^{\prime 2} e^{-g^\prime/2} \cos 2\pi
 q^\prime  \label{NPRG1}
 \\
 \beta_q &=& \frac{d q^\prime}{d\ln \beta}  = ~~~~~~~~~~ -
 \frac{D}{4\pi} g^{\prime 2} e^{-g^\prime/2} \sin 2\pi q^\prime 
 \label{NPRG2}
\end{eqnarray}
we obtain the renormalization group functions $\beta_{g,q} =
\beta_{g,q}(g^\prime,q^\prime)$ of the AES theory on the weak coupling side.
We have included the two loop correction~\cite{HZ}  in the perturbative part of Eq.~\eqref{NPRG1}.

Eqs~\eqref{NPRG1} and \eqref{NPRG2} are amongst the most important results of this investigation. The results clearly demonstrate that instantons are the fundamental topological objects of the AES theory that describe the cross-over behavior of the SET between the conducting phase at high temperatures and the 
Coulomb blockade phase that generally appears at much lower temperatures only. 

\section{The strong coupling problem, $g^\prime\ll 1$ \label{Sec:Strong}}
%
%
%
%
%
%
%
%
%

\subsection{Effective action for $\theta \approx \pi$}
For small values of the tunneling conductance $g$ we can simplify the
Hamiltonian of Eq.~\eqref{HamStart} near the degeneracy point 
$\theta = \pi$ or $q=1/2$ by employing a projection onto the
states with $Q=k(q)$ and $Q=k(q)+1$ of the isolated island.~\cite{Matveev} 
The projected Hamiltonian can be written as follows
\begin{gather}
 \mathcal{H} = \mathcal{H}_0 +  \mathcal{H}_c +
 \mathcal{H}_T^{(l)}+ \mathcal{H}_T^{(r)} \label{HamStartP}
\end{gather}
where
\begin{gather}
 \mathcal{H}_c = E_c (k-q)^2+\frac{\Delta}{2} - \Delta S^z
 ,\label{HamCoulP}
 \\ \mathcal{H}_T^{(s)} = \sum_{k\alpha}
t_{k\alpha}^{(s)} a^{(s)\dag}_{k} d_{\alpha} S^+ +
\textrm{h.c.}\end{gather}
Here, $\Delta=E_c(1-\theta/\pi)\geqslant 0$, $\bm{S}$ denotes the spin $s=1/2$ operators and $S^{\pm} = S^x\pm i S^y$.

A convenient representation is obtained by using Abrikosov's
two-component pseudo fermion fields $\bar{\psi}$ and
$\psi$.~\cite{Abrikosov,IzyumovSkryabin} After integration over
the electronic degrees of freedom one arrives at the following effective 
action to leading order in $1/N^{(s)}_{\rm ch}$
\begin{gather}
 \mathcal{S} = \beta E_c (k-q)^2 + \beta \frac{\Delta}{2}+\int_0^\beta
 d\tau \bar{\psi} \left (\partial_\tau -\eta +\frac{\Delta}{2}
 \sigma_z\right )\psi \notag \\ +\frac{g}{4}\int_0^\beta d\tau_1 d\tau_2
 \alpha(\tau_{12})[\bar{\psi}(\tau_1)\sigma_-\psi(\tau_1)]
 [\bar{\psi}(\tau_2)\sigma_+\psi(\tau_2)] .\label{Scrit}
\end{gather}
Here, $\sigma_j$ with $j=x,y,z$ stand for the Pauli matrices and
$\sigma_{\pm} = (\sigma_{x}\pm i\sigma_y)/2$. We have introduced
the chemical potential $\eta$ such that the limit $\eta\to
-\infty$ is taken at the end of all calculations. This procedure
ensures that only the physical states with pseudo fermion number
$N_{pf} = 1$ contribute to the quantities of physical interest.
Following the prescription~\cite{Abrikosov,IzyumovSkryabin}
\begin{equation}
 Z = \lim\limits_{\eta\to-\infty} \frac{\partial}{\partial e^{\beta
 \eta}}\, Z_{pf} \label{ZDef}
\end{equation}
we obtain the physical partition function $Z$ from the
pseudo fermion theory $Z_{pf}$. Similarly, we extract the physical
expectation $\langle \mathcal{O} \rangle$ according to
\begin{equation}
 \langle \mathcal{O} \rangle = \lim_{\eta\to-\infty}\left [
 \frac{Z_{pf}}{Z} \frac{\partial}{\partial e^{\beta\eta}} \,
 \langle \mathcal{O} \rangle_{pf}+\langle \mathcal{O}
 \rangle_{pf}\right ] .\label{ODef}
\end{equation}
The brackets $\langle\dots\rangle_{pf}$ denote the average with respect
to the theory of Eq.~\eqref{Scrit}.

In what follows we employ the effective action of Eq.~\eqref{Scrit} 
to investigate the phenomenon of macroscopic charge quantization 
as well as the renormalization behavior of the SET 
on the strong coupling side. Eq.~\eqref{Scrit} is similar to the $XY$ case of the Bose-Kondo model for spin $s=1/2$.~\cite{LarkinMelnikov,Si,Demler}
Notice that the spin operators $\bar{\psi}(\tau)\sigma_{\pm}\psi(\tau)$ 
in Eq.~\eqref{Scrit} are the same as the AES operators $\exp(\pm i \Phi(\tau))$ projected onto the states $Q=k$ and $Q=k+1$ of the isolated island. 
The projection onto the Hamiltonian of Eq.~\eqref{HamStartP} is justified
as long as $g \ll 1$, $|q-k-1/2|\ll 1$ and $\beta E_c\gg 1$.~\cite{GG3} 
\subsection{Leading logarithmic approximation}
In what follows we limit the analysis of Eq.~\eqref{Scrit} to the
so-called leading logarithmic approximation. This corresponds to
the one-loop renormalization group procedure of
Refs.~[\onlinecite{Si,Demler}].
%
%
\begin{figure}[t]
\includegraphics[width=55mm]{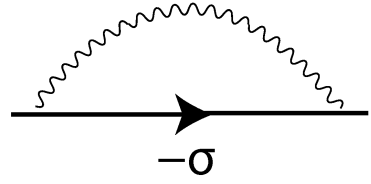}
\caption{The pseudofermion self-energy: solid line denotes
$G(i\epsilon_n)$ whereas wavy line stands for the interaction
$\alpha(i\omega_n)$ (see text).} \label{Fig:Diag1}
\end{figure}

%
\subsubsection{Pseudofermion Green function renormalization}
Using Eq.~\eqref{Scrit} we find the following expression for the
pseudofermion Green function for $g=0$
\begin{equation}
 G_{0\pm}^{-1}(i\epsilon_n) = i\epsilon_n +\eta \mp \frac{\Delta}{2},\label{G0bare}
\end{equation}
where $\epsilon_n = \pi T(2 n+1)$. The pseudofermion Green
function can be expressed in terms of the self-energy $\Sigma_\pm$
\begin{equation}
 G_{\pm}^{-1}(i\epsilon_n) = i\epsilon_n +\eta \mp \frac{\Delta}{2} -
 \Sigma_{\pm}(i\epsilon_n). \label{GFDef}
\end{equation}
It is convenient to parameterize the self-energy as follows
\begin{equation}
 \Sigma_{\pm}(i\epsilon_n) = (i\epsilon_n +\eta)[1-
 \gamma(i\epsilon_n)] \mp [1-\gamma_s(i\epsilon_n)] \frac{\Delta}{2} .
\end{equation}
The pseudofermion Green function now becomes
\begin{equation}
 G_{\pm}^{-1}(i\epsilon_n) = (i\epsilon_n +\eta)
 \gamma(i\epsilon_n) \mp \gamma_s(i\epsilon_n) \frac{\Delta}{2} .
\end{equation}
%
%
The leading logarithmic approximation corresponds to the simplest
diagram for the self-energy shown in Fig.~\ref{Fig:Diag1}. This
leads to the following equation
\begin{equation}
 \Sigma_{\pm}(i\epsilon_n) = -\frac{g T}{4\pi}\sum_{\omega_l}
 \frac{|\omega_l|}{i\omega_l+i\epsilon_n+\eta \pm \frac{\Delta}{2} -
 \Sigma_{\mp}(i\omega_l+i\epsilon_n)} \label{GenS}
\end{equation}
which has to be solved self consistently. Recall that there is no
renormalization of the interaction line (see
Fig.~\eqref{Fig:Diag1}) because of the absence of closed fermion
loops in the pseudofermion diagrammatic technique, i.e. their
contribution vanishes in the limit
$\eta\to-\infty$.~\cite{Abrikosov,IzyumovSkryabin}

With logarithmic accuracy we see that both $\gamma$ and $\gamma_s$
depend on the single variable $x = \ln \Lambda/\max\{\Delta
\gamma_s/\gamma,|i\epsilon_n+\eta|\}$ where $\Lambda$ is an
arbitrary high energy cut-off. Then from Eq.~\eqref{GenS} we
obtain
\begin{gather}
 \gamma(x) = 1 +\frac{g}{4\pi^2} \int_0^x\frac{d y}{\gamma(y)}, \\
 \gamma_s(x) = 1 -\frac{g}{4\pi^2} \int_0^x\frac{d
 y\,\gamma_s(y)}{\gamma^2(y)}
\end{gather}
and, hence,
\begin{equation}
 \gamma(x) = \gamma_s^{-1}(x) = \left (1+\frac{g}{2\pi^2} x \right
 )^{1/2} . \label{gammas}
\end{equation}
\subsubsection{The partition function and the average charge $Q$}
Using Eq.~\eqref{ZDef} we find
\begin{equation}
 Z = e^{\beta E_c (k-q)^2} e^{\beta\Delta/2}\lim_{\eta\to-\infty}
 e^{-\beta \eta} \sum_{\epsilon_n,\sigma=\pm} e^{i\epsilon_n 0^+}
 G_\sigma(i\epsilon_n).
\end{equation}
Given the vertex functions $\gamma(x)$ and $\gamma_s(x)$ it is now
trivial to evaluate the partition function
\begin{equation}
 Z=2 e^{\beta E_c (k-q)^2} e^{\beta\Delta/2} \cosh
 \frac{\beta \Delta^\prime}{2}.
\end{equation}
Here,
\begin{gather}
 \gamma =  \left (1+\frac{g}{2\pi^2} \ln
 \frac{\Lambda}{\max\{\Delta^\prime,T\}} \right )^{1/2},
\end{gather}
and $\Delta^\prime=\Delta/\gamma^2$ stands for the renormalized
energy gap between the ground state and first excited state.

The average charge on the island is expressed in terms of the
magnetization ${M}=\langle S_z \rangle$ of our spin model
\begin{equation}
 Q(T) = k+\frac{1}{2}  - {M} \label{QSCres-1}
\end{equation}
where
\begin{equation}
 \mathcal{M}(T) = \frac{1}{2 \gamma^2} \tanh
 \frac{\beta \Delta^\prime}{2} .\label{QSCres-2}
\end{equation}
Eq.~\eqref{QSCres-2} has originally been obtained in
Ref.~[\onlinecite{Schon}] using slightly different techniques.
Evaluating the result at $T= 0$ we find
\begin{equation} \label{Matveev}
 Q(T=0) = k+\frac{1}{2} - \frac{1}{1+\frac{g}{2\pi^2}
 \ln \frac{\Lambda}{\Delta^\prime} }
\end{equation}
which is the familiar result of Matveev.~\cite{Matveev} It does
not resemble the simple expression for the
averaged charge on an isolated island. This charge is, in fact, no
longer quantized when the tunneling conductance $g$ is finite, see
Section \ref{zero-T}.
%
\begin{figure}[t]
\includegraphics[width=55mm]{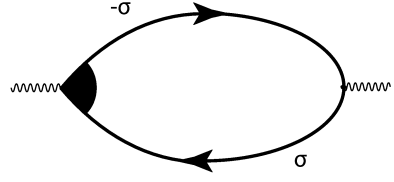}
\caption{The two-point correlation function: solid line denotes
$G(i\epsilon_n)$ whereas black triangle denotes the vertex
function $\Gamma(i\epsilon_n)$ (see text).} \label{Fig:Diag2}
\end{figure}

%
\subsubsection{The correlation function $D^R(\omega)$}
\begin{figure}[t]
\includegraphics[width=30mm]{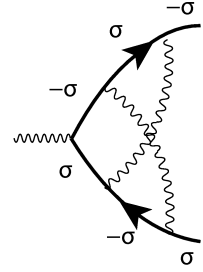}
\caption{The first non-trivial contribution to the vertex function
$\Gamma(i\epsilon_n)$ (see text).} \label{Fig:Diag3}
\end{figure}

The diagram for the two point correlation function $D(i\omega_n)$
is shown in Fig.~\ref{Fig:Diag2}. Because of the peculiar form of
the pseudofermion interaction which couples the $\sigma_-$ and
$\sigma_+$ it is readily seen that the lowest order contribution
to the vertex function $\Gamma(i\epsilon_n)$ is proportional to
$g^2 \ln (\Lambda/\max\{T,\Delta^\prime\})$, see
Fig.~\ref{Fig:Diag3}. Within the leading logarithmic approximation
one can therefore put $\Gamma(i\epsilon_n)=1$ and, hence,
\begin{gather}
 D(i\omega_n) = - \frac{1}{2\cosh\beta
 \Delta^\prime/2}\lim\limits_{\eta\to-\infty}
 \frac{\partial}{\partial e^{\beta \eta}}\, T \sum_{\epsilon_m}
 G_-(i\epsilon_m)\notag \\ \times G_+(i\epsilon_m+i\omega_n) =-
 \frac{\tanh \beta \Delta^\prime/2}{\gamma^2} \frac{1}{i\omega_n -
 \Delta^\prime} .\label{DMres}
\end{gather}
After analytic continuation to real frequencies we obtain
\begin{equation}
 D^R(\omega) = -\frac{\tanh \beta \Delta^\prime/2}{\gamma^2}
 \frac{1}{\omega- \Delta^\prime + i0^+} .\label{DRSC}
\end{equation}
%
%
%
%
\subsection{Physical observables\label{TDiss}}
Given Eq.~\eqref{DRSC}, it is possible to evaluate integrals in
Eqs.~\eqref{g-prime-def}, \eqref{theta-prime-def} (definitions I) and in
Eqs.~\eqref{gPrKor} and \eqref{qPrKor} (definitions II) and compute the response
functions in the strong coupling phase. For both definitions (I)
and (II) we obtain the same expression for the quasi particle
charge $q^\prime$
\begin{eqnarray}
 q^\prime &=& Q + \frac{1-\gamma^2}{2 \gamma^2}\tanh \frac{\beta
 \Delta^\prime}{2} \label{qPrSCres0} \\
 &=& k(q) + \frac{1}{e^{\beta \Delta^\prime}+1} . \label{qPrSCres}
\end{eqnarray}
However, different expressions for the response quantity
$g^\prime$ are obtained and the result is
\begin{eqnarray} g^\prime_{I} &=& \frac{g}{2 \gamma^2}
 \frac{\tanh \beta \Delta^\prime /2}{\beta \Delta^\prime /2}, \label{gPrSCres-0}\\
 g^\prime_{II} &=& \frac{g}{2 \gamma^2} \frac{\beta
 \Delta^\prime}{\sinh \beta \Delta^\prime} .\label{gPrSCres}
\end{eqnarray}
Eq.~\eqref{gPrSCres} coincides with the result found in
Ref.~[\onlinecite{Schon}] using a somewhat different approach. It
furthermore corresponds to the sequential tunneling approximation
of Ref.~[\onlinecite{KulikShekhter}].

There are several interesting conclusions that one can draw from
these findings. First of all, we see that as $T$ approaches
absolute zero Eqs \eqref{qPrSCres0} - \eqref{gPrSCres} are
independent of $g$ and precisely coincide with the results
obtained for the isolated island. Unlike Eq.\eqref{Matveev}, for
example, we find that the new quantity $q^\prime$ is robustly
quantized with infinitely sharp steps occurring when the external
charge $q$ passes through $k+1/2$.

However, Eqs \eqref{gPrSCres-0} and \eqref{gPrSCres} for the
response quantity $g^\prime$ do not unequivocally predict an
exponential dependence on $T$ when $\beta \Delta^\prime \gg 1$.
Moreover, as shown in Appendix~\ref{2Order}, the corrections to the quantities $g^\prime_{II}$ and $q^\prime_I$ to second order in $g$ do not demonstrate an exponential dependence on $T$ when $T$ vanishes. 

This means that the strong coupling expansion in powers of $g$
generally does not provide access to the Coulomb blockade phase
where the SET develops an energy gap. The validity of the leading
logarithmic approximation is therefore limited to the quantum
critical phase $\beta \Delta^\prime \lesssim 1$ which for our
purposes is the most significant regime of the SET.

This takes us to the most important part of this exercise which is
to employ Eqs \eqref{qPrSCres0} - \eqref{gPrSCres} in order to
extract the scaling behavior of the SET on the strong coupling
side. Since our physical observables are essentially defined for
finite size systems (i.e. finite $\beta$) they should in general
be distinguished from the ordinary thermodynamic quantities of the
quantum spin system that are normally being considered. Emerging
from Eqs \eqref{qPrSCres0} - \eqref{gPrSCres} there are two distinctly
different renormalization group schemes, to be discussed further
below, that provide complementary information on the quantum
system at zero temperature and finite temperatures respectively.
\subsubsection{RG at zero temperature \label{zero-T}}
Eqs \eqref{qPrSCres0} - \eqref{gPrSCres} clearly show that two
renormalizations are in general necessary to absorb the
ultraviolet or high energy singularity structure of our spin
system, i.e. one renormalization associated with the coupling
constant (tunneling conductance) $g$ and one associated with the
``magnetic field" (energy gap) $\Delta$. From the expressions at
zero $T$
\begin{equation}
 \frac{g}{\gamma^2} = \frac{g}{1+\frac{g}{2\pi^2} \ln
 \frac{\Lambda}{\Delta^\prime}} ~,~~ \Delta^\prime = \frac{\Delta}{\gamma^2} =
 \frac{\Delta}{1+\frac{g}{2\pi^2} \ln
 \frac{\Lambda}{\Delta^\prime}}
\end{equation}
we obtain the following renormalization group $\beta$ and $\gamma$
functions to one loop order
\begin{equation}
 \beta_g = \frac{d g}{d \ln \Lambda} = \frac{g^2}{2\pi^2}~,~~
 \gamma_\Delta = \frac{d \ln \Delta}{d \ln \Lambda} =
 \frac{g}{2\pi^2}.
\end{equation}
Employing the method of characteristics one can cast the
thermodynamic quantities of the quantum spin system at $T=0$ in a
general scaling form. For example, the magnetization $M$ with
varying ``magnetic field" $\Delta$ can be expressed as follows
\begin{equation}
 M (\Delta) = M_0 f( \Delta M_0 \xi) .
\end{equation}
Here, the functions $M_0$ and $\xi$ with varying $g$ are
determined by the $\beta$ and $\gamma$ functions according to
\begin{eqnarray}
 \left( \Lambda \frac{\partial}{\partial \Lambda} + \beta_g
 \frac{\partial}{\partial g} \right) \xi &=& 0 \\
 \left( \gamma_\Delta + \beta_g
 \frac{\partial}{\partial g} \right) M_0 &=& 0 .
\end{eqnarray}
One finds, for example, that the characteristic time scale $\xi$
of the SET is given by
\begin{equation} \label{xi-strong}
 \xi = \Lambda^{-1} e^{-2\pi^2 /g}
\end{equation}
which has the same meaning as the weak coupling expression of Eq.
\eqref{xi-weak}. The quantity $M_0$ and the scaling function
$f(X)$ within the one loop approximation are given by
\begin{equation}
 M_0 = 1/2g ~,~~ f(X) = 2\pi^2 \ln^{-1} X^{-1} .
\end{equation}
%
The result essentially tells us that the spontaneous magnetization
only exists for the theory with $g=0$ but it vanishes for any
finite value of $g$. In terms of the AES model this means that the
averaged charge $Q$ on the island is no longer quantized when
finite values of the tunneling conductance are taken into account.
\subsubsection{RG at finite temperatures}
We next specialize to the physical observables at finite
temperature. Since the expressions of Eqs \eqref{qPrSCres0} -
\eqref{gPrSCres} are universal for $\beta\Delta^\prime \ll 1$
\begin{gather}
 g^\prime(T) \backsimeq \frac{g}{2}\left (1+\frac{g}{2\pi^2}\ln
 \beta
 \Lambda\right )^{-1},\label{W1} \\
 q^\prime (T) \backsimeq k + \frac{1}{2} - \frac{\beta \Delta}{4} \left
 (1+\frac{g}{2\pi^2}\ln \beta \Lambda\right )^{-1} \label{W2}
\end{gather}
we immediately obtain the finite temperature $\beta$ functions
along the critical lines $q^\prime = k + 1/2$  for $g^\prime \ll 1$
according to
\begin{eqnarray}
 \beta_g &=& \frac{d g^\prime}{d\ln \beta} = -\frac{g^{\prime 2}}{\pi^2} 
 \label{SCb}  \\
 \beta_q &=& \frac{d q^\prime}{d\ln \beta} = \left(q^\prime -k-\frac{1}{2} \right) 
 \left (1-\frac{g^\prime}{\pi^2}\right ) .
 \label{SCb-1} 
\end{eqnarray}
These results should be compared with Eq.~\eqref{strong-RG}
obtained for an isolated island. We see that the critical fixed 
point of an isolated island is the critical fixed point of the AES 
theory as a whole with the SET conductance $g^\prime$ now playing 
the role of a {\em marginally irrelevant} scaling variable. 

\begin{figure}[t]
\includegraphics[width=65mm]{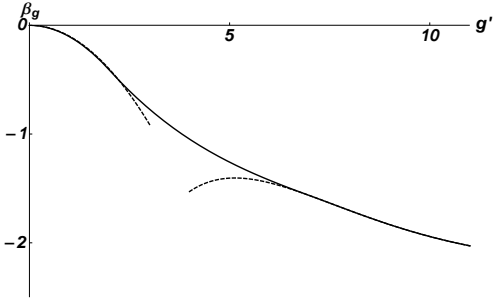}
\caption{The function $\beta_g$ with varying $g^\prime$ along the critical line 
$q^\prime = k+1/2$. An interpolation between the weak and strong coupling 
branches has been drawn as a guide for the eye, see text.}
\label{betag}
\end{figure}

\begin{figure}[t]
\includegraphics[width=65mm]{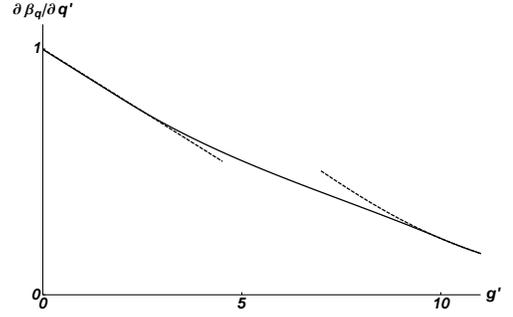}
\caption{ The function $\partial \beta_q/\partial q^\prime$ with varying $g^\prime$ along the critical line $q^\prime = k+1/2$. An interpolation between the weak and strong coupling branches has been drawn as a guide for the eye, see text.}
\label{betaq}
\end{figure}

Next we compare Eqs \eqref{SCb} and \eqref{SCb-1} with the the weak coupling
results of Eqs ~\eqref{NPRG1} and \eqref{NPRG2}. In Figs \ref{betag} and \ref{betaq}
we plot the functions $\beta_g$ and $\partial \beta_q /\partial q$ along the critical line $q^\prime =k+1/2$. A simple interpolation between the weak and 
strong coupling branches indicates that both these functions decrease 
monotonically as $g^\prime$ increases. 

Finally, it is not difficult to understand why the Coulomb blockade phase of the SET
is beyond the scope of the present investigation. For example, given the fact that the theory with $q^\prime \approx k $ and $g^\prime \ll 1$ develops an 
energy gap then one generally expects (see also Eq. \eqref{stable-strong})
\begin{eqnarray}
 \beta_g &=& {g^{\prime}} \ln g^\prime \\
 \beta_q &=& \left(q^\prime -k\right) \ln |q^\prime -k| \label{SCb-2}
\end{eqnarray}
which cannot be obtained using ordinary perturbation theory in $g^\prime$.
%
%
%
%
%
\section{Summary and conclusions\label{Sec:End}}
\begin{figure}[t]
\includegraphics[height=65mm]{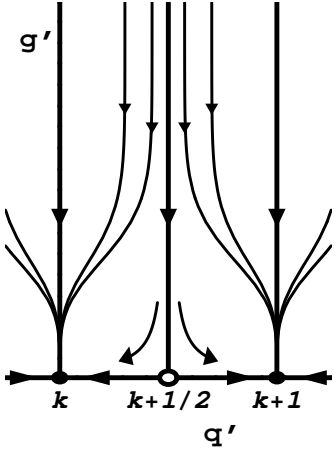}
\caption{Unified scaling diagram of the Coulomb blockade in terms
of the SET conductance $g^\prime$ and the $q^\prime$. The arrows
indicate the scaling toward $T = 0$ (see text).} \label{RGFIG}
\end{figure}

To summarize the results of this investigation we have sketched, 
in Fig.~\ref{RGFIG}, a unifying scaling diagram of the SET in 
the $g^\prime$ - $q^\prime$ plane. This diagram is based on 
the strong coupling results of Eqs ~\eqref{strong-RG} , \eqref{SCb} 
and \eqref{SCb-1} and the weak coupling results of Eqs ~\eqref{NPRG1} 
and \eqref{NPRG2}.

The universal features of this diagram are the quantum critical fixed points 
located at $q^\prime = k + 1/2$, $g^\prime =0$, and the stable fixed points at
$q^\prime = k$, $g^\prime =0$ that describe the ``macroscopic charge 
quantization" of the SET. The results are in accordance with the concept 
of super universality that has previously been proposed in the context of the 
quantum Hall effect.~\cite{PruiskenBurmistrov2}

We have established the relation between the quantity $g^\prime$ and the
ordinary SET conductance $G$ that one normally obtains from linear response 
theory. The quantity $q^\prime$ is new and can similarly be expressed in terms 
of the antisymmetric current-current correlation function. 

The quantization of $q^\prime$ is an interesting and important challenge for experimental research on single electron devices. There are, however, other ways of experimentally probing the quasi particle charge $q^\prime$ of the SET. In Section \ref{quantum-crit} below we will summarize the quantum critical properties of $q^\prime$ and point out how they are directly measurable in the experiment.

We conclude this paper with Section \ref{q-prime-change} below where we discuss in some detail the physical mechanism that is responsible for changing the the quasi particle charge $q^\prime$ of the SET as $q$ passes through the critical point. 
%
%
\subsection{Quantum criticality \label{quantum-crit}}
Eq. \eqref{W1} is the maximum value of $g^\prime (T)$ as one varies the value of 
$q$. This maximum value vanishes logarithmically in $T$ according to
\begin{eqnarray}
 g^\prime_{max} (T) &=& ~ \frac{g}{2\gamma^2} ~ =
 \frac{g}{2}\left ( 1+\frac{g}{2\pi^2}\ln {\beta \Lambda}\right )^{-1} \nonumber \\
 &=& {\pi^2}{\ln^{-1} \left( {\beta}/{\xi} \right)} \ll 1 .\label{g-max}
\end{eqnarray}
Similarly, Eq. \eqref{W2} determines the maximum slope of the quasi particle charge $q^\prime(T)$ with varying $q$. This slope diverges according to
\begin{eqnarray}
 \left[ \frac{\partial q^\prime (T)}{\partial q} \right]_{max}
 &=& \frac{\beta E_c} {2}\left ( 1+\frac{g}{2\pi^2}\ln
 {\beta \Lambda}\right )^{-1} \nonumber \\
 &=& \frac{\pi^2}{g} {(\beta E_c)} {\ln^{-1} \left( {\beta}/{\xi} \right)}.
 \label{Deltaq}
\end{eqnarray}
The inverse of this quantity is a measure for the width $\Delta q$ of the
transition. This width vanishes as $T$ goes to zero. 

Notice that Eq. \eqref{Deltaq} is completely analogous to what happens at the plateau transitions in the quantum Hall regime. In that case we have 
\begin{equation}
 \left[ \frac{\partial R_H}{\partial B} \right]_{max} \propto \beta^{\kappa} ,
\end{equation}
i.e. the maximum slope of the Hall conductance $R_H$ with
varying magnetic field $B$ diverges algebraically in $\beta$. The best experimental value of the critical exponent $\kappa$ equals $0.42$. Similar to the experiments on the quantum Hall effect, one may consider higher order derivatives of $g^\prime$ and $q^\prime$ with
respect to $q$.~\cite{QHE_Exp} These quantities diverge even faster with higher powers of
$\beta$.

The quantity $\partial q^\prime/\partial q$ can directly be measured in the experiment since it determines the {\em renormalized} gate capacitance of the SET $C_g^\prime = C_g \partial q^\prime/\partial q$. As was shown recently, the 
{\em rate of energy dissipation} ($P$) in the SET due to a low frequency gate voltage $V_g(t) = V_g + U_\omega \cos\omega t$ is given by~\cite{RBA}
\begin{equation} \label{P}
 P \propto \omega^2 |U_\omega|^2 ~ \frac{C_g^{\prime 2}}{g^\prime} . 
\end{equation}
Therefore, the maximum in $P$ with varying values of $V_g$ diverges according to
\begin{equation} \label{P-max}
 P_{max} \propto \left[ \frac{C_g^{\prime 2}}{g^\prime} \right]_{max} = \frac{\pi^2  
 C_g^2 }{g^2} (\beta E_c)^2\ln^{-1} (\beta/\xi) .
\end{equation}

Finally, it should be mentioned that the critical behavior of the
SET is likely to change when the effective number of channels 
$N_{\rm ch}^{(l,r)}$ between
the island and the reservoirs are finite rather than infinite.~\cite{Nazarov,Cond} Even though we expect that our theory of
physical observables remains unchanged, Matveev~\cite{Matveev}
has argued that the critical behavior of the SET can be mapped
onto the $N$-channel Kondo model.~\cite{Kondo} This would mean
that the transition at $q^\prime = k+1/2$ becomes a second order
one with a finite critical value of $g^\prime$ thus closely
resembling the more complicated physics of the quantum Hall
effect.~\cite{PruiskenBurmistrov2} Progress along these lines will
be reported elsewhere.~\cite{Elsewhere}
\subsection{Quantization of $q^\prime$ \label{q-prime-change}}
We have seen that the Thouless criterion for the Coulomb blockade breaks down at points $q=k+1/2$ where the energy gap $\Delta^\prime$ vanishes. 
To understand how the critical features of the SET permit a change in $q^\prime$ 
one must think in terms of a dynamical process where a unit of external
charge is added to the system at (imaginary) time $0$ and
removed at $\tau >0$. This process is described
by the two-point correlation function $D(\tau)$ given by
\begin{equation}
 D (\tau) = \langle e^{i\Phi(0) -i\Phi(\tau)} \rangle . \label{2-p-corr}
\end{equation}
Following Eq. \eqref{g-prime-def}, the tunneling through the SET involves the sum over all processes $D(\tau)$ according to
\begin{equation} \label{tunn}
 g^\prime = g T \int_0^\beta d\tau D(\tau) .
\end{equation}
From Eq.~\eqref{DMres} we obtain the following expression
valid at $T=0$ when $q$ approaches $k+1/2$ from below
\begin{equation}
 D (\tau) = \gamma^{-2} \Theta(\tau)
 e^{-\tau \Delta^\prime}. \label{charge-time}
\end{equation}
This general result includes the isolated island except
that the AES operators are now renormalized ($\gamma \ne 1$) and the energy gap $\Delta$ is replaced by the renormalized value $\Delta^\prime$. 

Let us first assume that $D(\tau)$ denotes the correlation of an isolated island. Eq. \eqref{tunn} then stands for a semi-classical picture of the SET where the island and reservoirs are essentially disconnected. Since the expectation value $\langle \tau \rangle$ is finite for $q < k + 1/2$
\begin{equation}
 \langle \tau \rangle = \frac{\int_0^\infty d\tau \tau D(\tau)}{\int_0^\infty d\tau  
 D(\tau)} = \frac{1}{\Delta}
\end{equation}
it is impossible that the tunneling processes described by Eq. \eqref{tunn} alter the static charge $Q$ on the island. However, as one approaches the critical point then the expectation $\langle \tau \rangle$ diverges. It is thus possible that when $q$ passes through $k+1/2$, a unit of charge stays behind on the island. This extra charge is precisely what lowers the energy of the island, i.e. it permits the energy to jump from one parabolic branch $E_c (q-k)^2$ to the next $E_c (q-k-1)^2$, see Fig.~\ref{Cusps}.

From the expression for $q^\prime$ in Eq. \eqref{theta-prime-def} it is clear
that this semi classical picture of the SET gets dramatically complicated when the tunneling conductance $g$ is finite. In particular, the second term proportional to $g$ is Eq. \eqref{theta-prime-def} clearly indicates that the quantization of $q^\prime$ goes hand in hand with strong charge fluctuations between the island and the reservoirs. Nevertheless, the {\em mechanism} for changing the quasi particle charge $q^\prime$ of the SET remains essentially the same. This mechanism solely involves a vanishing energy gap $\Delta^\prime$. The only difference with the semi classical picture is that the AES operators $e^{\pm i\Phi}$ in Eq. \eqref{2-p-corr} generally stand for the {\em quasi-particle} operators of the SET, rather than those of ordinary electrons in an isolated island.

Let us next consider the tunneling process in some more detail. We are interested, first of all, in the energy difference $\delta E$ between the states $|q+1 \rangle$ and $|q \rangle$ of the SET. Here, $|q+1 \rangle$
is formally defined as follows
\begin{equation}
 |q+1 \rangle = \lim_{\tau_0 \rightarrow \beta} |q(\tau) \rangle
\end{equation}
where $q(\tau) = q+1$ for $0<\tau<\tau_0$ and $q(\tau) = q$ for $\tau_0<\tau<\beta$.
After elementary algebra we obtain~\cite{Elsewhere}
\begin{equation}\label{delta-E}
 \delta E = \frac{\Delta^\prime}{1+e^{-\beta \Delta^\prime}} + T \frac{1}{\gamma^2}  
 \frac{d \gamma^2 }{d\ln T}.
\end{equation}
Since $\delta E \sim \Delta^\prime$ at low temperatures ($T \leqslant \Delta^\prime$) we conclude that the transition from $|q \rangle$ to $|q+1 \rangle$ is energetically unfavorable. 

Next, the rates for electron tunneling from reservoir to island ($\Gamma_{01}$) and backward ($\Gamma_{10})$ are computed to be~\cite{KulikShekhter,RBA}
\begin{equation}
 \Gamma_{01/10} = \frac{g \Delta^\prime}{4\pi \gamma^2}\left ( \coth\frac{ 
 \Delta^\prime}{2T} \pm 1\right ) .
\end{equation}
As long as the energy gap $\Delta^\prime$ is finite, the energy difference 
$\delta E$ in Eq. \eqref{delta-E} and the tunneling rates $\Gamma_{01/10}$ are not related to one another in any obvious manner. However, at the critical point $\Delta=0$ we find
\begin{equation}
\delta E = \frac{1}{\beta \ln \beta / \xi} 
~~,~~ \Gamma_{01} = \Gamma_{10} = \pi \delta E .
\end{equation}
Hence, the energy difference between the states $|q+1 \rangle$ and 
$|q \rangle $ determines the time the electron resides on the island.
It is therefore possible that the tunneling processes alter the 
static charge $q^\prime$ of the SET as $q$ passes through $k+1/2$.

\begin{acknowledgements}

The authors are grateful to A. Abanov, O. Astafiev, A. Finkelstein, A. Ioselevich, 
A. Lebedev and Yu. Makhlin for stimulating discussions. The research
was funded in part by the Dutch National Science Foundations
\textit{NWO} and \textit{FOM}, the Russian Ministry of Education
and Science, \textit{CRDF}, the Council for Grant of the President
of Russian Federation (Grant No. MK-125.2009.2), \textit{RFBR} (Grant No. 09-02-92474-MHKC and No. 07-02-00998), the Dynasty Foundation and \textit{RAS} Programs ``Quantum Physics of Condensed Matter'' and 
``Fundamentals of nanotechnology and nanomaterials''.

\end{acknowledgements}

\appendix

\section{Analytic continuation of $K(i\omega_n)$ \label{App:AnalCont}}

In this appendix we perform the various steps that take us from
$K(i\omega_n)$, Eq.~\eqref{KMdef}, to the expression for
$K^R(\omega)$ in Eq.~\eqref{KRdef}. Starting from the correlation
function
\begin{gather}
 K(i\omega_n) = -\frac{g}{4\pi} T \sum_{\omega_m}
 |\omega_m+\omega_n| D(i\omega_m)
\end{gather}
we employ the following relations
 \begin{gather}
 |\omega_m| = \int \frac{d \epsilon}{\pi}\frac{i\omega_m}{\epsilon+i\omega_m}, \\
 D(i\omega_m) = \int \frac{d \epsilon}{\pi} \frac{\Im D^R(\epsilon)}{\epsilon-i\omega_m} \label{A3}
\end{gather}
and obtain the following expression
\begin{gather}
 K(i\omega_n) = \frac{g}{4\pi} \int \frac{d \epsilon_1 d \epsilon_2}{\pi^2}
 \frac{\epsilon_2 \Im D^R(\epsilon_1)}{\epsilon_1+\epsilon_2+i\omega_n}\notag \\
 \hspace{1.5cm}\times T\sum_{\omega_m} \left [
 \frac{1}{\epsilon_1-i\omega_m}+\frac{1}{\epsilon_2+i\omega_m+i\omega_n}\right ]
 .
\end{gather}
Evaluating the sum over $\omega_m$ we find
\begin{gather}
 K(i\omega_n) =  \frac{g}{4\pi} \int \frac{d \epsilon_1 d \epsilon_2}{\pi^2}
 \frac{\epsilon_2 \Im D^R(\epsilon_1)}{\epsilon_1-\epsilon_2+i\omega_n}\notag \\
 \hspace{2cm}\times [n_b(\epsilon_2)-n_b(\epsilon_1)].\label{K_A1}
\end{gather}
The analytic continuation to real frequencies is now trivial and
we directly obtain Eq.~\eqref{KRdef}.

\section{Useful identities \label{Iden}}
The identities obtained in this Section will be of use in Appendix
\ref{2Order}. First, we consider the derivative of the average
charge $Q$ with respect to $g$ which can be obtained as follows
\begin{eqnarray}
 \frac{\partial Q}{\partial g} &=&
 \frac{\partial^2 \ln Z_{pf}}{\partial g \partial \Delta}
 \nonumber \\
 &=& \frac{1}{4\pi}
 \frac{\partial}{\partial \Delta} T \sum_{\omega_n} |\omega_n|
 D(i\omega_n) .\label{QApp}
\end{eqnarray}
With the help of Eq.~\eqref{A3}, it is convenient to rewrite Eq.~\eqref{QApp} in the following
manner
\begin{equation}
 \frac{\partial Q}{\partial g} = \frac{\partial}{\partial \Delta}
 \int \frac{d\epsilon}{2\pi^2} Y(\epsilon) \Im D^R(\epsilon)
 ,\label{QIden}
\end{equation}
where
\begin{equation}
 Y(\epsilon) = T\sum_{\omega_n>0} \frac{\omega_n
 \epsilon}{\omega_n^2+\epsilon^2} .
\end{equation}
A second useful identity for the expression appearing in
Eq.~\eqref{thetaPrBack} is given by
\begin{equation}
 \Re \frac{\partial K^R(\omega)}{\partial \omega}
 = - \frac{g}{2\pi^2} \int d\epsilon
 \Im D^R(\epsilon) \frac{\partial Y(\epsilon)}{\partial \epsilon} \label{ReKIden}
\end{equation}
where the limit $\omega \rightarrow 0$ is understood.

\section{Evaluation of $q^\prime$ to second order in $g$\label{2Order}}
Based on Eqs \eqref{QIden} and \eqref{ReKIden} we evaluate, in
this Appendix, the expression for $q^\prime$ in Eq.
\eqref{thetaPrBack} to second order in $g$. We start from the
two-point correlation function $D(i\omega_n)$ which to first order
in $g$ is given by
\begin{eqnarray}
 D(i\omega_n) = &-& \frac{\tanh(\beta \Delta/2)}{i\omega_n-\Delta}
 \left [ 1 - \frac{g}{\pi} \frac{\beta Y(\Delta)
 }{\sinh(\beta\Delta)}\right ] \notag \\
 &+& \frac{g}{\pi}
 \tanh(\beta\Delta/2) \frac{\partial}{\partial \Delta}
 \left( \frac{Y(\Delta)}{i\omega_n-\Delta} \right) \notag \\
 &-&  \frac{g}{4\pi} \frac{|\omega_n|}{(i\omega_n-\Delta)^2} .
 \label{DApp1}
\end{eqnarray}
%
%
%
This result can be written in the following form
\begin{equation}
 D(i\omega_n)= -\frac{\tanh (\beta\Delta_1^\prime/2)}{\gamma^2_1}
 \frac{1}{i\omega_n-\Delta^\prime_1} - \frac{g}{4\pi} \frac{|\omega_n|}{(i\omega_n-\Delta)^2} .
 \label{Dapp2}
\end{equation}
where the renormalized energy gap and the renormalization factor are given as 
\begin{gather}
 \Delta_1^\prime = \Delta - \frac{g}{\pi} Y(\Delta), \qquad
 \frac{1}{\gamma_1^2} = \frac{\partial
 \Delta_1^\prime}{\partial\Delta} .
\end{gather}
Eq.~\eqref{Dapp2} implies the following expression for the retarded function:
\begin{eqnarray}
 D^R(\epsilon)= &-&\frac{\tanh (\beta\Delta_1^\prime/2)}{\gamma^2_1}
 \frac{1}{\epsilon-\Delta^\prime_1+i0^+} \notag \\
 &+& \frac{g}{4\pi} \frac{i\epsilon}{(\epsilon-\Delta+i0^+)^2} .
 \label{Dapp2a}
\end{eqnarray}

By inserting the result~\eqref{Dapp2a} for $D^R (\epsilon)$ in
Eq.~\eqref{ReKIden} we find after elementary algebra
\begin{gather}
 \Re \frac{\partial K_1^R(\omega)}{\partial \omega}
 = - \frac{g}{2\pi}
 \frac{\tanh(\beta\Delta_1^\prime/2)}{\gamma_1^2} \frac{\partial
 Y(\Delta_1) }{\partial\Delta_1} \notag \\
 + \frac{g^2}{32\pi^2} \frac{\partial}{\partial\Delta}
 \left (\Delta \frac{\partial}{\partial\Delta}\Delta \coth \frac{\beta\Delta}{2} \right )
 \label{KRapp}
\end{gather}
where the limit $\omega\to 0$ is understood.
Next, by expanding Eq. \eqref{KRapp} to the second order in $g$ we
finally obtain
\begin{eqnarray}
 \Re \frac{\partial K_1^R(\omega)}{\partial \omega}
 = &-&\frac{g}{2\pi} \tanh \frac{\beta\Delta}{2} \notag \\
 &\times &\Biggl
 [ \frac{\partial Y}{\partial \Delta} - \frac{g}{2\pi}
 \frac{\partial^2 Y^2}{\partial \Delta^2} 
 - \frac{g}{2\pi} \frac{\beta}{\sinh \beta\Delta} \frac{\partial
 Y^2}{\partial \Delta} \Biggr ] \notag \\
 &+& \frac{g^2}{32\pi^2}
 \frac{\partial}{\partial\Delta}\left (\Delta
 \frac{\partial}{\partial\Delta}\Delta \coth \frac{\beta\Delta}{2} \right ).
 \label{KRapp2}
\end{eqnarray}

We proceed by inserting the result for $D^R (\epsilon)$ in the
expression of Eq. ~\eqref{QIden} and find
\begin{gather}
 \frac{\partial Q}{\partial g} = \frac{1}{2\pi}
 \frac{\partial}{\partial\Delta}\left [
 \frac{\tanh(\beta\Delta_1^\prime/2)}{\gamma_1^2}
 Y(\Delta_1^\prime) \right ] \label{QRapp}\\
 -\frac{g}{32\pi^2} \frac{\partial^2}{\partial\Delta^2}
 \left (\Delta^2 \coth\frac{\beta\Delta}{2}\right ) .
\end{gather}
Up to second in $g$ the expression for the averaged charge $Q$
therefore becomes
\begin{gather}
Q= k(q)+\frac{1}{1+e^{\beta\Delta}} + \frac{g}{2\pi}
 \frac{\partial}{\partial\Delta}\Biggl [ \tanh
 \frac{\beta\Delta}{2} \Bigl ( Y - \frac{g}{2\pi} \frac{\partial
 Y^2}{\partial \Delta} \notag \\
 - \frac{g}{2\pi} \frac{\beta Y^2}{\sinh \beta\Delta}  \Bigr ) \Biggr
 ]-\frac{g^2}{64\pi^2} \frac{\partial^2}{\partial\Delta^2}
 \left (\Delta^2 \coth\frac{\beta\Delta}{2}\right )
 .\label{Qapp2}
\end{gather}
Finally, collecting Eqs.~\eqref{KRapp2} and \eqref{Qapp2} together
we find the total result for $q^\prime$ as follows
\begin{widetext}
\begin{gather}
 q^\prime = k(q) + \frac{1}{1+e^{\beta\Delta}} + \frac{g}{2\pi} \left(
 Y \frac{\partial}{\partial\Delta}
 - \frac{g}{2\pi} \frac{\partial Y^2}{\partial\Delta}
 \frac{\partial}{\partial\Delta} -
 \frac{g}{2\pi} Y^2 \frac{\partial^2}{\partial\Delta^2} \right)
 \tanh \frac{\beta\Delta}{2}
 +\frac{g^2}{64\pi^2} \Delta\frac{\partial^2}{\partial\Delta^2}
 \left (\Delta \coth\frac{\beta\Delta}{2}\right ) . 
\end{gather}
\end{widetext}
The result can be written in a slightly more compact fashion
according to
\begin{gather}
 q^\prime = k(q)+\frac{1}{1+e^{\beta\Delta^\prime_2}}+
 \frac{g^2}{64\pi^2} \Delta\frac{\partial^2}{\partial\Delta^2}
 \left (\Delta \coth\frac{\beta\Delta}{2}\right ) .\label{Dapp10}
\end{gather}
Here,  
\begin{equation}
 \Delta_2^\prime = \Delta - (g/\pi) Y(\Delta_1^\prime)
\end{equation}
represents the second order in $g$ expression for the renormalized gap. 

Similarly, with the help of Eqs.~\eqref{DApp1} and~\eqref{Dapp2a} from Eqs.~\eqref{gPrBack}, \eqref{g-prime-def}, and \eqref{theta-prime-def} one can compute the other response parameters to the second order in $g$. The results can be summarized as follows:
\begin{eqnarray}
g^\prime_{I}&=& \frac{g}{2\gamma_1^2} \frac{\tanh \beta \Delta^\prime_1/2}{\beta\Delta^\prime_1/2},\\
q_{I}^\prime &=& k(q)+\frac{1}{1+e^{\beta\Delta^\prime_2}}-
 \frac{g^2\Delta}{64\pi^2} \frac{\partial^2}{\partial\Delta^2}
 \left (\Delta \coth\frac{\beta\Delta}{2}\right ) \notag \\
 && +\frac{g^2}{8\pi^3}\beta 
 \Delta^\prime_1 \tanh \frac{\beta\Delta^\prime_1}{2} 
 \Imag \psi^\prime \left (1+\frac{i\beta\Delta}{2\pi}\right ),
 \end{eqnarray}
and
\begin{eqnarray}
g^\prime_{II}&=& \frac{g}{2\gamma_1^2} \frac{\beta \Delta^\prime_1}{\sinh\beta\Delta^\prime_1}\notag \\
&& -\frac{g^2}{4\pi^2} \Bigl [ 1 +\frac{\partial}{\partial\Delta} \frac{\beta\Delta^2}{2\pi} \Imag \psi^\prime \left (1+\frac{i\beta\Delta}{2\pi}\right )\Bigr ] ,\\
q_{II}^\prime &=& k(q)+\frac{1}{1+e^{\beta\Delta^\prime_2}}+
 \frac{g^2\Delta}{64\pi^2} \frac{\partial^2}{\partial\Delta^2}
 \left (\Delta \coth\frac{\beta\Delta}{2}\right ) .
 \end{eqnarray}

These results are different from those obtained in the leading logarithmic approximation. In some cases (i.e. $g^\prime_{II}$ and $q^\prime_I$) the corrections in $g$ no longer predict an exponential dependence on $T$ in the limit where $T$ goes to zero. This clearly shows that the expansion to lowest orders in $g$ does not provide access to the Coulomb blockade phase of the SET.

\section{Evaluation of $K^R(\omega)$\label{genKR}}

In this Appendix we present the results of explicit computations of the response function $K^R(\omega)$. These results can be used, first of all, as an independent check on the results of Eqs. ~\eqref{Pert}, \eqref{gPrInst}, \eqref{qPrInst}, \eqref{qPrSCres} and \eqref{gPrSCres}. Secondly, they show how the analytic continuation of $K(i\omega_n)$ to real frequencies works in explicit computations.

\subsection{Weak coupling regime $g^\prime\gg 1$}

Based on Eqs ~\eqref{DRPert} and \eqref{DRInst} we obtain the
following expression from Eq.~\eqref{KRdef}
\begin{widetext}
\begin{eqnarray}
 K^R(\omega) &=& \frac{i\omega g}{4\pi} \left [ 1- \frac{2}{g}\ln
 \frac{e g E_c}{2\pi^2 T} +\frac{2}{g}\psi \left
 (1-\frac{i\omega}{2\pi T}\right ) \right ]
 -  \sum_{n=1}^{n_\textrm{max}} \frac{\frac{g E_c}{2\pi^2}}{n+\frac{g
 E_c}{2\pi^2 T}}\notag \\
 && - \frac{g^3 E_c}{2\pi^2} e^{-g/2} \left\{ e^{i 2\pi q} \Bigl [\psi \left
 (1\right )-\psi \left (1-\frac{i\omega}{2\pi T}\right ) \Bigr]
 + \cos 2\pi q \sum_{n=2}^{n_\textrm{max}}
 \frac{1}{n}  \right\}.
\end{eqnarray}
%
Here, the cuttoff $n_\textrm{max}$ appears due to the fact that, as usual, we use the low-frequency part of the kernel $\alpha(\tau)$ only. If one takes the proper  expression for it into account then one finds 
$n_\textrm{max} \sim E_F/T$ where $E_F$ denotes the Fermi energy.

\subsection{Strong coupling regime $g^\prime\ll 1$}

Given Eq.~\eqref{DRSC} we can write
\begin{gather}
 K^R(\omega) =  -\frac{g \tanh
 \frac{\beta\Delta^\prime}{2}}{4\pi^2\gamma^2}
 \int d \epsilon\, \epsilon\, \frac{n_b(\epsilon)-n_b(\Delta^\prime)}
 {\epsilon -\Delta^\prime-\omega-i0^+} .
\end{gather}
Hence,
%
\begin{eqnarray}
 \Im K^R(\omega) &=& \frac{g \tanh
 \frac{\beta\Delta^\prime}{2}}{4\pi\gamma^2}(\omega+\Delta^\prime)
 [n_b(\Delta^\prime)-n_b(\omega+\Delta^\prime)] \\
 \Re K^R(\omega) &=& -\frac{g \tanh
 \frac{\beta\Delta^\prime}{2}}{2\pi\gamma^2}
 Y(\omega+\Delta^\prime) \notag \\
 &=& \frac{g \tanh \frac{\beta\Delta^\prime}{2}}{4\pi^2\gamma^2}
 (\Delta^\prime+\omega)\Re\Bigl [ \psi
 \left (1-\frac{i(\Delta^\prime+\omega)}{2\pi T}\right )
 -\psi\left (1-\frac{i\Delta^\prime}{2\pi T}\right )
 -\frac{2\pi}{\Delta^\prime}Y(\Delta^\prime)\Bigr ]
\end{eqnarray}
%
By using the following representation of Bose-Einstein function in the sum over Matsubara frequencies
\begin{equation}
 n_b(\omega)= \frac{1}{\pi}\Im \psi\left (1+i \frac{\omega}{2\pi T}\right ) -\frac{1}{2}+\frac{T}{\omega}
 \end{equation}
we finally obtain
%
\begin{equation}
 K^R(\omega) = \frac{g \tanh
 \frac{\beta\Delta^\prime}{2}}{4\pi^2\gamma^2} (\omega+\Delta^\prime) \Bigl [
 \psi\left (1-i \frac{\omega+\Delta^\prime}{2\pi T}\right )
 -\psi\left (1-i \frac{\Delta^\prime}{2\pi T}\right )
 -\frac{2\pi}{\Delta^\prime}Y(\Delta^\prime)+\frac{i T\omega}
 {\Delta^\prime(\omega+\Delta^\prime)} \Bigr ].
\end{equation}
\end{widetext}

\end{document}